\documentclass[reprint,amsmath,amssymb,aps,pra,longbibliography]{revtex4-2}

\usepackage{graphicx}
\usepackage{epsfig}
\usepackage{epstopdf}
\usepackage{subfigure,color}
\usepackage{dcolumn}
\usepackage{bm}
\usepackage{lineno}
\usepackage{amsmath, amssymb}
\usepackage{accents}
\usepackage{lipsum}
\usepackage{multirow}
\usepackage[usestackEOL]{stackengine}
\usepackage{tabularx}
\usepackage{makecell}
\usepackage{xcolor}
\usepackage{dcolumn}
\usepackage{bm}
\usepackage{booktabs}
\def\e{\begin{equation}}
\def\f{\end{equation}}
\def\_#1{{\bf #1}}
\def\/#1{_{\rm #1}}
\def\.{\cdot}

\begin{document}

\title{Coherently time-varying metasurfaces}

\author{M.~H.~Mostafa$^{1}$} 
\email{mohamed.mostafa@aalto.fi}
\author{A.~D{\'i}az-Rubio$^{1, 2}$} 
\author{M.~S.~Mirmoosa$^{1}$} 
\author{S.~A.~Tretyakov$^{1}$} 

\affiliation{$^{1}$Department~of~Electronics~and~Nanoengineering, Aalto~University, P.O.~Box~15500, FI-00076~Aalto, Finland\\ 
$^{2}$Nanophotonics Technology~Center, Universitat Politècnica de València, Valencia 46022, Spain}
	
	
\begin{abstract} 

Known coherent metasurfaces control interference of waves of a given frequency with other coherent waves at the same frequency, either illuminating from a different direction or created as intermodulation products.  In this paper, we introduce a class of metasurfaces that are modulated in time coherently with the illuminating radiation. Importantly, such modulation opens a possibility to control reflection, absorption, and transmission at multiple frequencies, including illuminations by two or more incoherent waves. 
In particular, we study dynamic resistive layers and show how to use them to design thin multi-frequency perfect absorbers that overcome the bandwidth limit for static linear absorbers. Furthermore, we demonstrate possibilities of remote tuning of the absorption level. We hope that this work opens up novel avenues in wave engineering using coherent modulation of metasurface parameters. 
\end{abstract}
	
\maketitle
	

\section{Introduction} 
	
Wave interference is the basis of numerous devices such as phased-array antennas~\cite{mailloux2017phased}, interferometers~\cite{interferometrybasics}, coherent absorbers~\cite{Chong2010,radi2015}, 
and Bragg reflectors. A classical example of the possibilities offered by interference phenomena in the design of practical devices is the Salisbury screen~\cite{salisbury}. In that device, full absorption of incident power in a lossy sheet of negligible thickness is realized by placing a mirror at the quarter-wave distance, creating destructive interference of reflected waves. Similar concept is applied to one class of 
coherent perfect absorbers that are based on coherent illumination of two sides of a sheet using a splitter or an independent phase-locked generator~\cite{baranov2017coherent,li2015broadband,fang2015controlling}, realizing destructive interference of reflected and transmitted waves required to ensure perfect absorption. Recently, this notion of coherent illumination of thin sheets was extended to surface-inhomogeneous coherent metasurfaces capable of locally controlling the interference of the illuminating waves along the surface~\cite{cuesta2020coherent}. Another more recent example of possibilities offered by wave interference is perfect absorption in complex scattering  and disordered media~\cite{Chen2020, Pichler}. It was shown that by inducing purposeful perturbations in random  medium’s  disorder,  it is possible to obtain perfect absorption conditions~\cite{Imani2020}. In all these works, e.g.~\cite{baranov2017coherent,li2015broadband,fang2015controlling,cuesta2020coherent,Chen2020, Pichler, Imani2020}, there is an important assumption that the systems are temporally stationary (time-invariant), and the interference phenomena are controlled only by illuminating waves.

An alternative approach to engineer interference phenomena is to exploit the frequency mixing produced in temporally nonstationary (time-varying) systems~\cite{tien1958Parametric}. For instance, one can consider classical parametric amplifiers where periodical modulation of a reactive element makes it possible to amplify signals passing through it~\cite{cullen1958}. In this case, to allow interaction between the input signal at a frequency $\omega_{\rm s}$ and the modulation products, the dynamic system should be coherently pumped at $\omega_{\rm m}=2\omega_{\rm s}$. The same operational principle has been applied for enhancing wireless transfer of power and information~\cite{jayathurathnage2020time} and for designing time-modulated metasurfaces with exotic properties such as nonreciprocity~\cite{wang2020nonreciprocity}. Similar to the static coherent systems that require precise synchronization of the incident waves, parametric systems relying on interference between the input signal and the modulation products also require synchronization.



\begin{figure}[t!]
\centerline
{\includegraphics[width=1 \columnwidth]{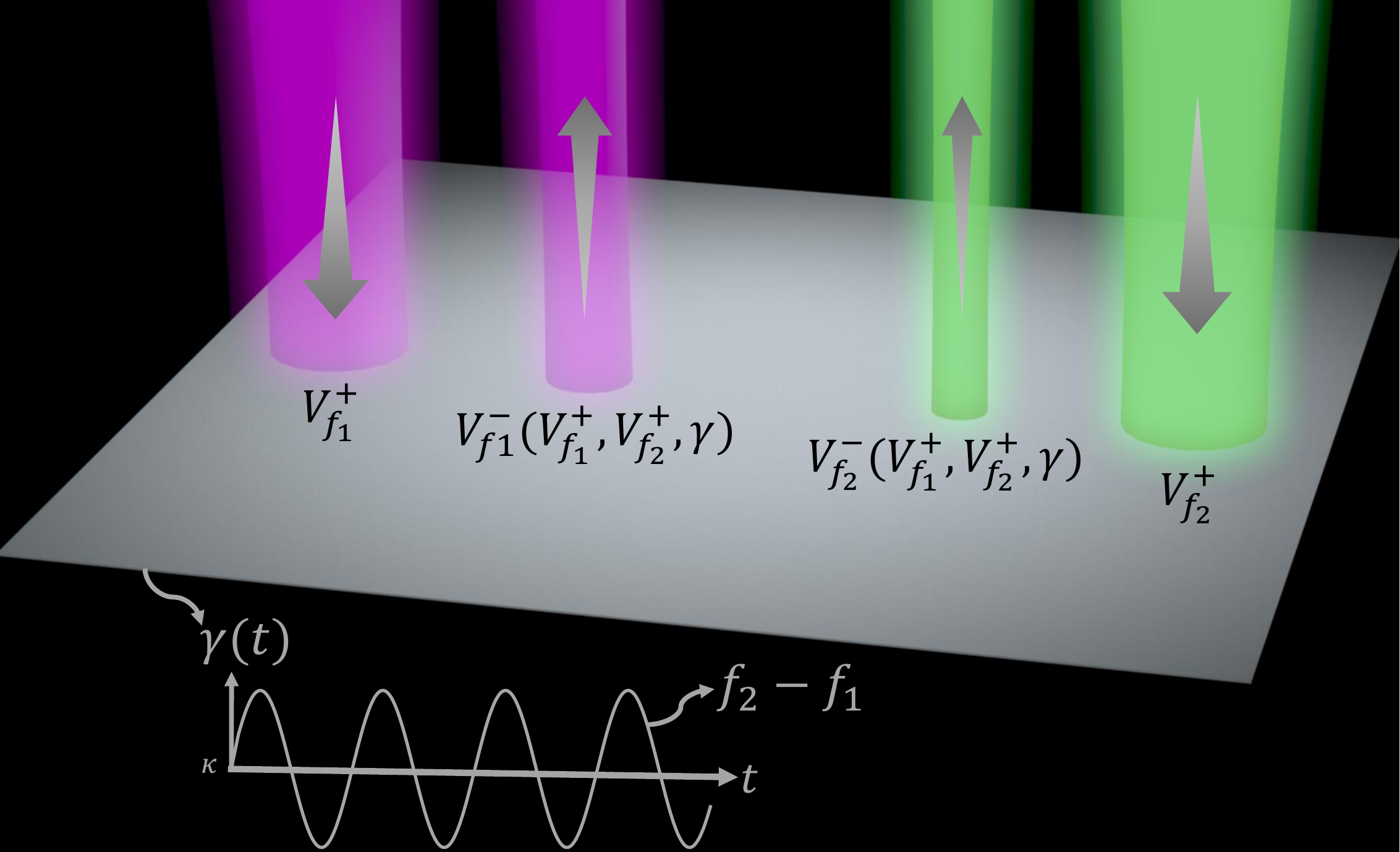}} 
\caption{Conceptual view for coherently time-varying boundary under multi-frequency illumination.} 
\label{fig0}
\end{figure}

Known coherent devices control interference of waves of a given frequency with other coherent waves at the same frequency, either illuminating from a different direction or created as intermodulation products. In this paper, we introduce coherently time-varying metasurfaces that are modulated at the beating frequency or frequencies of two or more incident waves. 
A conceptual schematic for such a metasurface illuminated by two waves at different frequencies is shown in Fig.~\ref{fig0}.
We show that coherent interactions of modulated metasurface with the frequency harmonics of the incident radiation can be employed to fully tailor the frequency response of the metasurface at multiple frequencies simultaneously.
In particular, we present a controllable thin metasurface that fully absorbs incident waves at several frequencies.


We start our analysis with a general study of dynamic resistive boundaries under multi-frequency illumination. In contrast to the well-studied time-varying reactive boundaries or metasurfaces (e.g.,~Refs.~\cite{HSTGM,AGSTMM,XWSTMMAT,XWSTM}) and reactive elements (e.g.,~Ref.~\cite{tien1958Parametric} and Refs.~\cite{macdonald1961exact,mirmoosa2019time,elnaggar2020}), time-varying resistive boundaries and elements have not been sufficiently studied, and there are only a few works about them~\cite{Tucker1963circuits,Peterson1927impedance,Glucksman1949,Manley1956,Peterson1939equivalent,Tucker1960rectifier}. We emphasise that, while temporal modulation has been intensively used recently as a technique to realize many applications (such as nonreciprocity~\cite{FAN2009,SOUNAS2017}, one-way beam splitting~\cite{TARAVATI2019}, power combining~\cite{XWSTM}, frequency conversion and generation of higher-order frequency harmonics~\cite{GRBIC2020FT,SALARY2019,PRXFT}, parametric amplification~\cite{FLEURY2018, Pendry2019Luminal}, enhancing wireless power transfer~\cite{jayathurathnage2020time}, control of scattering and radiation~\cite{Engheta2019aiming, Hayran:21, Ptitcyn2019, M2020PRA, ArXivPTVP}, and so forth), much less attention has been devoted to the temporal modulation of losses in electromagnetic systems despite the possibility to offer intriguing features and promising applications. In this paper, we uncover that coherently time-varying lossy boundaries under 
multi-frequency illuminations are able to exhibit tunable virtual reactive response.  Subsequently, we show how to efficiently use this characteristic to fully control the frequency response of the metasurface. In addition, we propose a realistic topology for the design of a multifrequency perfect absorber that overcomes the bandwidth limit of linear static absorbers. For any time-invariant and passive metasurface absorber there is a trade-off between the absorber  thickness and the bandwidth. In particular, the Rozanov limit~\cite{Rozanov} defines the upper bound for the bandwidth to thickness ratio for linear static Dällenbach screen absorbers. The limit elucidates that perfect absorption can be realized only for monochromatic input, at a single frequency, and the same conclusion applies to any single-resonance metasurface absorber. Here, we show that by introducing proper coherent time modulation of a resistive sheet, it is possible to create a linear thin absorber that perfectly absorbs multiple frequencies simultaneously, overcoming this bandwidth limitation (see Appendix~\ref{appa}). Compared to other means for enhancing absorption (reducing reflections) of metasurfaces which are based on spreading reflections over wide frequency band and maintaining reflections under certain level~\cite{ Chambers2005, caloz2021Camouflaging, Huanan2021}, the proposed approach ensures increasing the power dissipated in the metasurface, which makes reflection negligible over the whole frequency spectrum. Finally, we show that it is also possible to remotely tune the level of absorption.

It is important to note that throughout our analysis, we consider modulation functions such that the effective resistance of the resistive sheet or boundary remains positive at all times. Thus, the system under study does not exhibit gain, and it remains lossy and unconditionally stable, unlike the case of time-varying reactive elements. Furthermore, we consider slow temporal modulation where the modulation frequency is much smaller than the input frequencies. This approach is much more practical than the fast modulation conventionally used in most parametric devices.

The paper is organized as follows. In Section~\ref{cirbt}, we theoretically study dynamic resistive boundaries under multi-frequency illumination. In Section~\ref{abcs}, we describe our proposed structure for a multifrequency perfect absorber. Finally, in Section~\ref{concl}, we conclude the paper. 


	
\section{Dynamic resistive boundary under multi-frequency illumination}
\label{cirbt}

\begin{figure*}
  \centering
  \begin{minipage}[t]{1\textwidth}
  \subfigure[]{\includegraphics[width=.32\linewidth]{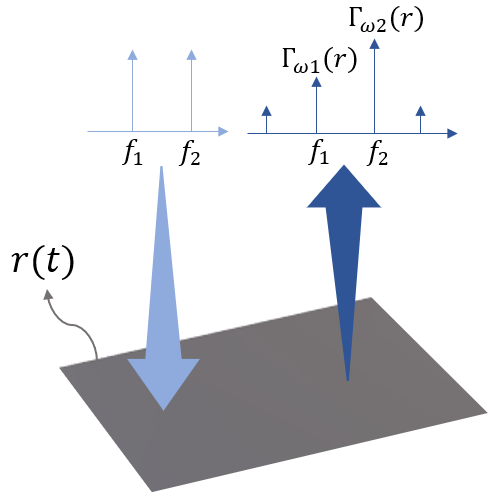} \label{fig1a}}
  \hspace{6em}%
  \subfigure[]{\includegraphics[width=0.23\linewidth]{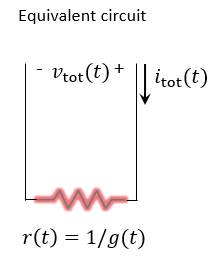} \label{fig1b}}
  \hspace{6em}%
  \subfigure[]{\includegraphics[width=0.18\linewidth]{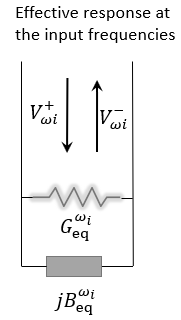} \label{fig1c}}
  \end{minipage}
  \begin{minipage}[t]{1\textwidth}
  \subfigure[]{\includegraphics[width=0.27\linewidth]{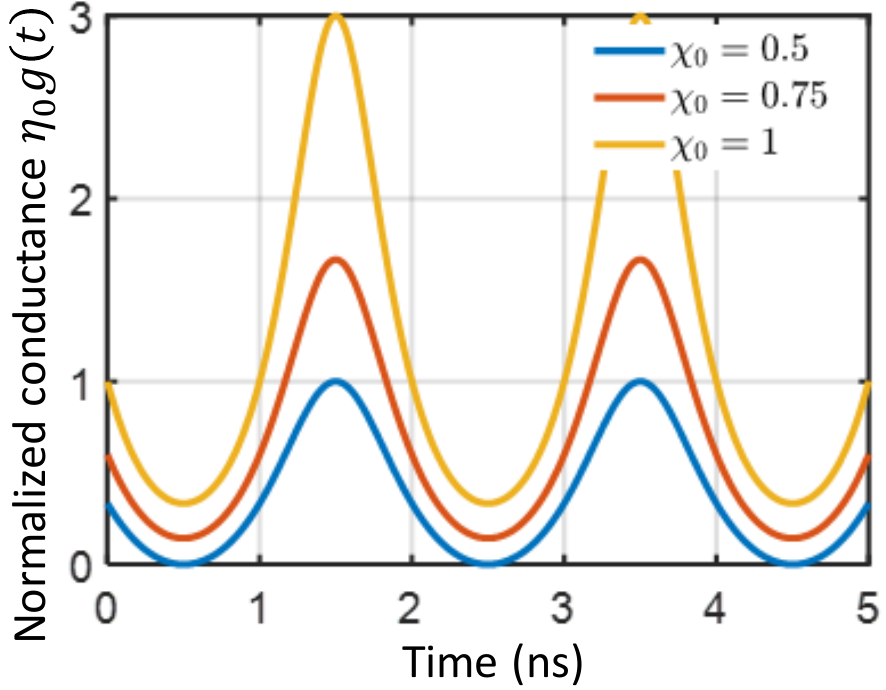} \label{fig1d}}
  \subfigure[]{\includegraphics[width=0.32\linewidth]{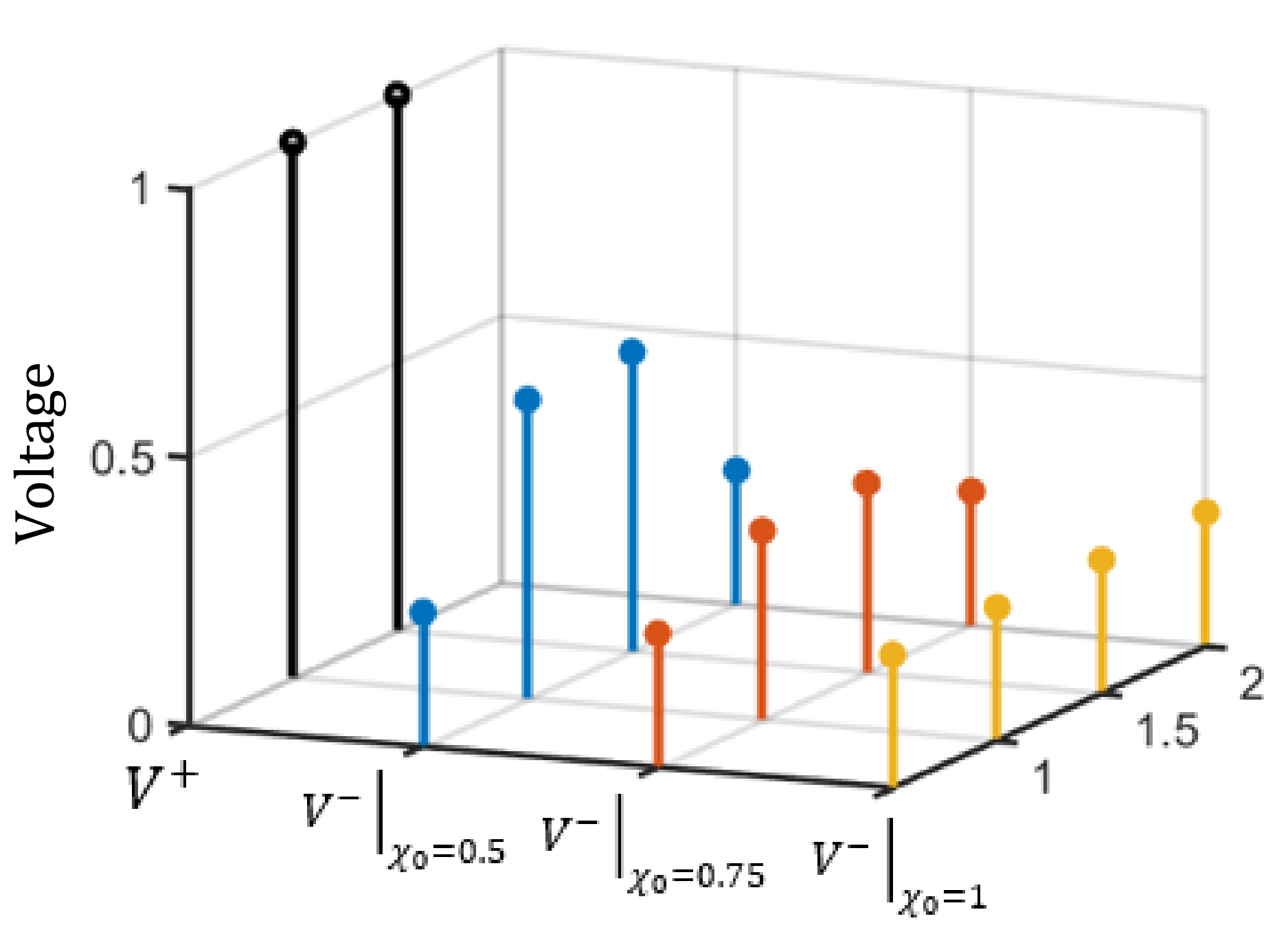} \label{fig1e}}
   \subfigure[]{\includegraphics[width=0.32\linewidth]{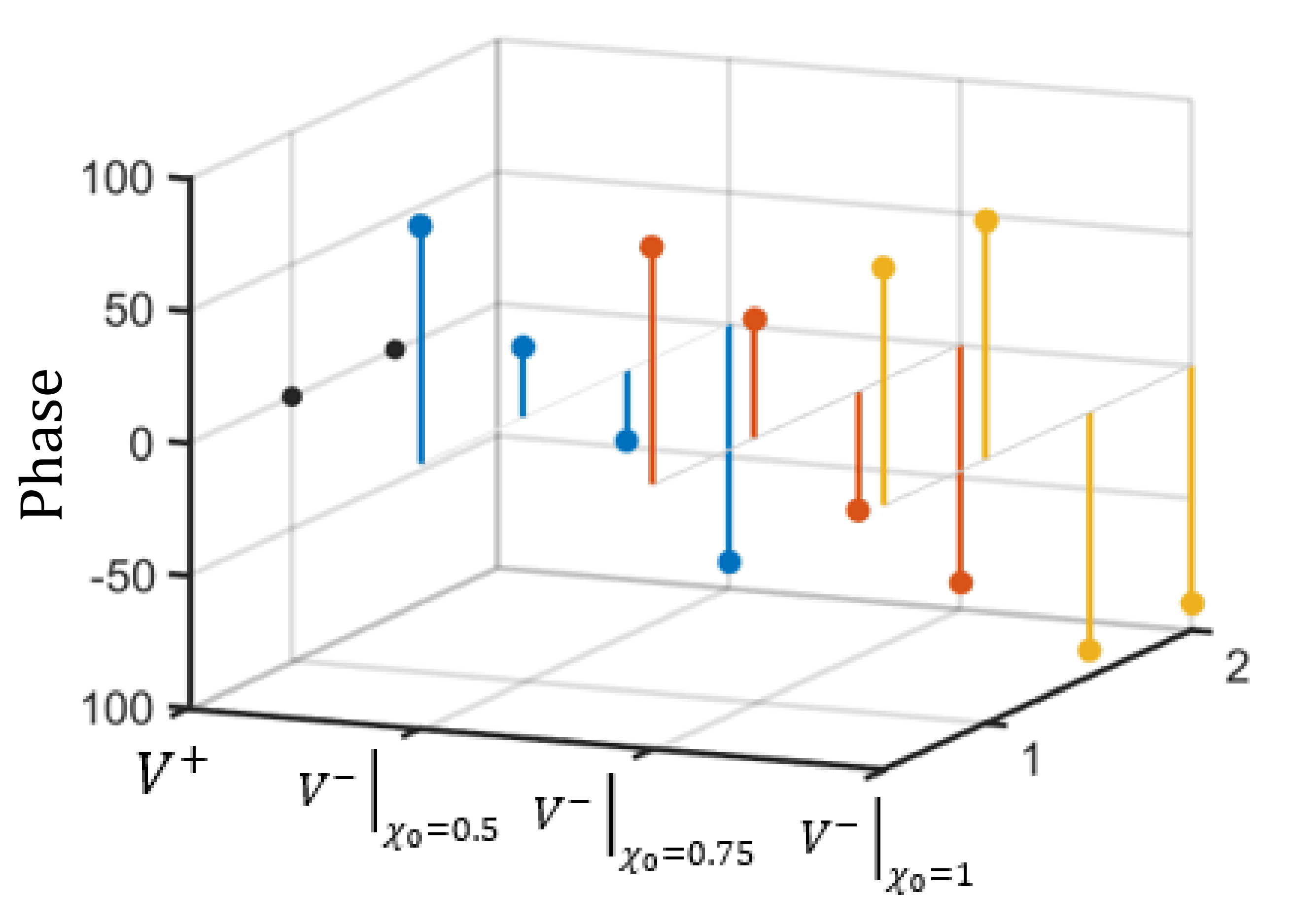} \label{fig1f}}
  \end{minipage}
  \caption{Multi-frequency illumination of a dynamic resistive boundary. (a) Schematic representation of the temporally modulated resistive boundary. (b) Equivalent circuit model of the proposed structure at normal incidence. (c) Effective response of the boundary at the input frequencies. (d)--(f) Study of the scattering properties when $f_1=1$ GHz, $f_2=1.5$ GHz, $a_1=a_2=1$~V, $\phi_1=\phi_2=0$, $\chi_{\rm m}=0.5$, and $\phi_{\rm m}=\pi/2$ for different values of $\chi_{\rm 0}$.}
  \label{fig:Fig1}
\end{figure*}

We start the analysis by considering a dispersionless flat resistive boundary that extends over the $xy$-plane and whose properties change in time, as it is illustrated in Fig.~\ref{fig1a}. The electromagnetic properties of the boundary can be modeled by the effective conductance $g(t)=1/r(t)$ in which $r(t)$ denotes the effective resistance. To simplify the mathematical derivations, we analyze the metasurface using a transmission-line model that is shown in Fig.~\ref{fig1b}. Based on this model, the tangential components of the total electric and magnetic fields on
the surface of the boundary are analogous to the total voltage $v_{\rm tot}$ and
electric current $i_{\rm tot}$ in the circuit ($E_{\rm tot} \xrightarrow{} v_{\rm tot}$ and $H_{\rm tot} \xrightarrow{} i_{\rm tot}$). Note that we assume that the electromagnetic fields do not penetrate behind the boundary, and, therefore, the resistive boundary modeled by a time-modulated resistor is terminating the transmission line.

At the boundary, the total voltage and current are written as combinations of the incident and reflected voltages  $ v_{\rm tot}(t)=v^+(t)+v^-(t)$ and $i_{\rm tot}(t)=[v^+(t)-v^-(t)]/\eta_0$, where $v^+$ and $v^-$ represent the incident and reflected voltages, respectively, and $\eta_0$ is the intrinsic impedance of the background medium (assumed to be free space). Also, the boundary condition at the termination reads $i_{\rm tot}=g(t)v_{\rm tot}$. By applying these relations, we find the waves reflected by the dynamic resistive boundary as  
\begin{equation}
  v^-(t)=\left[1- \frac{2g(t)\eta_0}{1+g(t)\eta_0}\right] v^+(t)=\gamma(t) v^+(t).
  \label{eq:reflectionco}
\end{equation}
In this expression, we define the parameter $\gamma(t)$ as the instantaneous reflection coefficient at the boundary. It is clear that by properly time modulating the conductance (or resistance) of the boundary, we can engineer the instantaneous reflection coefficient, and, consequently, the scattering properties at different frequencies. 

In what follows, we will study how temporal variation of conductance can be used to control the interference phenomena when the boundary is simultaneously illuminated by two plane waves at different frequencies. 
In this scenario, the input voltages can be written as
\begin{equation}
v^+(t)=a_{1}\cos{(\omega_{1}t+\phi_{\rm 1})}+a_{2}\cos{(\omega_{2}t+\phi_{\rm 2})},
\end{equation}
where $a_{1,2}$, $\omega_{1,2}$, and $\phi_{1,2}$ are the amplitudes, frequencies, and phases of the input harmonics, respectively. Let us write the instantaneous reflection coefficient as $\gamma(t)=1-\chi(t)$ and consider the temporal variation of the form $\chi(t)=\chi_0+\chi_{\rm m}\cos{(\omega_{\rm m}t+\phi_{\rm m})}$ in order to simplify the analysis. Now, by using these definitions, the reflected voltages are expressed as 
\begin{equation}
v^-(t)= (1-\chi_0)v^+(t)-\chi_{\rm m} \cos{(\omega_{\rm m}t+\phi_{\rm m})}v^+(t). 
\label{eq:rx_boundary}
\end{equation}
According to Eq.~(\ref{eq:reflectionco}), we simply conclude that the function $\chi(t)$ has maximum and minimum values of $2$ and $0$, respectively, for positive values of $g(t)$. Thus, since we consider a lossy resistive boundary at every moment in time [$g(t)>0$], the modulation parameters are constrained to
\begin{equation} 
\chi_0+|\chi_{\rm m}|<2 , \quad \chi_0-|\chi_{\rm m}|>0.
\label{conditions}
\end{equation}

From the expression of the reflected voltage in Eq.~(\ref{eq:rx_boundary}), we see that in order to enhance the interaction between harmonics, the modulation frequency should satisfy $\omega_{\rm m}=\omega_2-\omega_{\rm 1}$. In this case, the signal is decomposed to four reflected harmonics: $v^-(t)=v^-_1(t)+v^-_2(t) +v^-_3(t) +v^-_4(t) $ that are given by 
\begin{equation}
\begin{split}
&v^-_1(t)=\cr
&(1-\chi_0)a_{1}\cos{(\omega_{1}t+\phi_{\rm 1})}-\frac{a_2}{2}\chi_{\rm m}\cos{(\omega_{1}t+\phi_{\rm 2}-\phi_{\rm m})},\cr
&v^-_2(t)=\cr
&(1-\chi_0)a_{2}\cos{(\omega_{2}t+\phi_{\rm 2})}-\frac{a_1}{2}\chi_{\rm m}\cos{(\omega_{2}t+\phi_{\rm m}+\phi_{\rm 1})},\cr
&v^-_3(t) =-\frac{a_1}{2}\chi_{\rm m}\cos{\Big[(\omega_{2}-2\omega_1)t+\phi_{\rm m}-\phi_{\rm 1}\Big]},\cr
&v^-_4(t) =-\frac{a_2}{2}\chi_{\rm m}\cos{\Big[(2\omega_{2}-\omega_1)t+\phi_{\rm m}+\phi_{\rm 2}\Big]}.
\end{split}
\end{equation}
Here, it is worth noting the difference with the conventional parametric systems where by modulating at the double frequency of the input signal, we control the interference produced at the input frequency. In the above expressions, the response at each input frequency depends on the amplitudes and phases of both input harmonics and on the modulation parameters, opening more degrees of freedom in engineering the desired response. In addition, we use slow modulation in which the modulation frequency is much smaller than the input frequencies. Hence, employing this approach is more practical than the double-frequency modulation.

To evaluate the effect produced by multi-frequency illuminations of dynamic boundaries, one can analyze the scattered fields at different frequencies. The complex amplitudes of the scattered waves are written as 
\begin{equation}
\begin{split}
&V^-_{\omega_1}= (1-\chi_0)a_{1}e^{j\phi_{\rm 1}} -\frac{a_2}{2}\chi_{\rm m}e^{j(\phi_2-\phi_{\rm m})},\\
&V^-_{\omega_2}= (1-\chi_0)a_{2}e^{j\phi_{\rm 2}} -\frac{a_1}{2}\chi_{\rm m}e^{j(\phi_{\rm m}+\phi_{\rm 1})},\\
&V^-_{\omega_2-2\omega_1}= -\frac{a_1}{2}\chi_{\rm m}e^{j(\phi_{\rm m}-\phi_{\rm 1})},\\ 
&V^-_{2\omega_2-\omega_1}= -\frac{a_2}{2}\chi_{\rm m}e^{j(\phi_{\rm m}+\phi_{\rm 2})}.
\end{split}
\label{eq:parasitic}
\end{equation}
Figure~\ref{fig:Fig1} presents the results of a study of the scattering properties of a metasurface when $f_1=1$~GHz, $f_2=1.5$~GHz, $a_1=a_2=1$~V, $\phi_1=\phi_2=0$, $f_{\rm m}=0.5$~GHz, $\chi_{\rm m}=0.5$, and $\phi_{\rm{m}}=\pi/2$ for different values of $\chi_{\rm 0}$. The temporal variation of the conductance for different values of $\chi_{\rm 0}$ is shown in Fig.~\ref{fig1d}, and the amplitudes and phases of the reflected waves are plotted in Fig.~\ref{fig1e} and Fig.~\ref{fig1f}, respectively. While the boundary is purely resistive, the reflected waves are not in phase with the incident waves due to the non-zero modulation phase. Thus, the phase of the reflected waves can be tuned by properly designing the modulation phase. In addition, we see that the parameter $\chi_{\rm 0}$ controls the difference between the amplitudes of the harmonics at the input frequencies and the parasitic harmonics. Based on Eq.~\eqref{eq:parasitic}, we observe that the amplitudes of the parasitic harmonics are controlled by varying the amplitude of the modulation function $\chi_{\rm m}$. This feature will play an important role in the design of practical devices where the presence of parasitic harmonics is undesired.  

\begin{figure}[t!]
\centering
\subfigure[]{\includegraphics[width=0.7\linewidth]{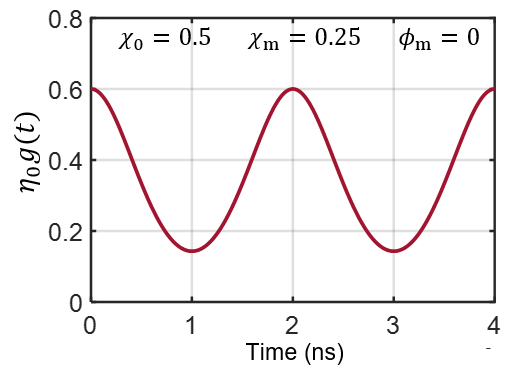} \label{fig4a}}
\subfigure[]{\includegraphics[width=0.7\linewidth]{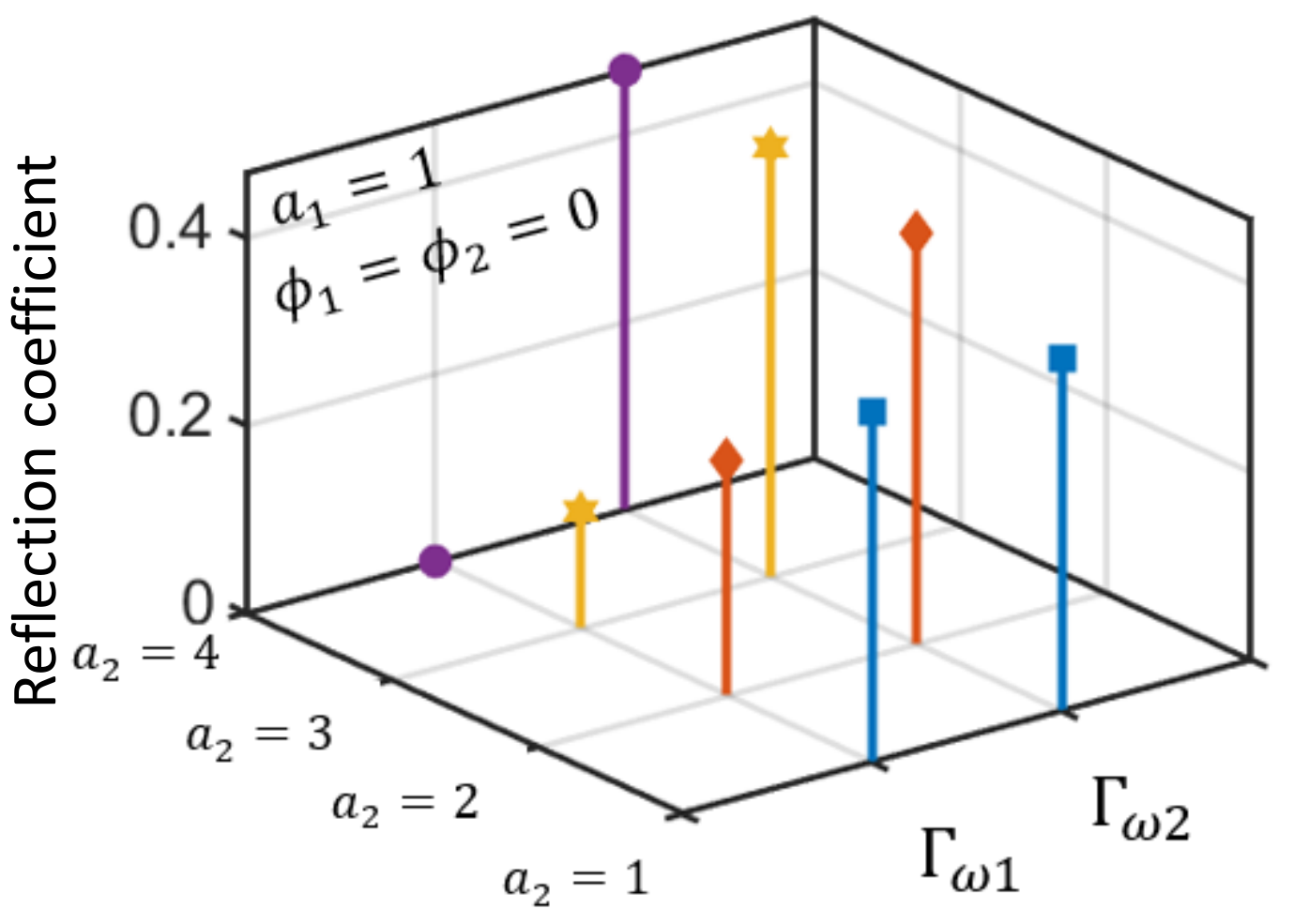}\label{fig4b}}
\caption{Control of the reflection coefficient by a dynamic resistive boundary. (a) Time variation of the normalized conductance, where $f_{\rm m}=0.5$~GHz. (b) Scattering properties of the metasurface at the input frequencies. } 
\label{fig:fig4}
\end{figure}

It is also important to notice that in this example, the amplitudes of the reflected waves at the input frequencies are equal because we have assumed that $a_1=a_2$. However, the amplitudes can be engineered to produce asymmetric reflection or even cancel reflection at one of the frequencies while allowing reflection at the other. 
The response of the boundary at the input frequencies can be analysed by using a complex reflection coefficient, defined as the ratio between the complex amplitude of the reflected and incident waves at each input frequency: $V_{\omega_i}^-=\Gamma_{\omega_i}V_{\omega_i}^+$. The expressions for these reflection coefficients read
\begin{equation}
\begin{split}
&\Gamma_{\omega_1}= (1-\chi_0) -\frac{a_2}{2a_{1}}\chi_{\rm m}e^{j(\phi_2-\phi_{\rm m}-\phi_{\rm 1})},\\
&\Gamma_{\omega_2}= (1-\chi_0) -\frac{a_1}{2a_{2}}\chi_{\rm m}e^{-j(\phi_2-\phi_{\rm m}-\phi_{\rm 1})}.
\end{split}
\end{equation}
As it is clear, the ratio between the amplitude of the input harmonics, the modulation properties, and the relation between the phases allow us to control not only the magnitude of the reflection coefficients but also the corresponding phases. Figure~\ref{fig:fig4} demonstrates how by varying the ratio between the amplitudes of the input harmonics, the reflection coefficients can be controlled. If the ratio is not unity, the reflection becomes asymmetric. We see that the effect of increasing $a_2$ is negligible on reflection at frequency $\omega_2$, but it is quite strong on reflection at frequency $\omega_1$. This asymmetric response is used later to reduce the parasitic harmonics generated due to the modulation, while inducing a strong effect at the input frequencies. 

For a static resistive boundary, the reflected waves are in phase with the incident waves. However, when the resistive boundary is modulated in time, we find that the phase of the reflected waves depends on the modulation phase. As a result, by designing the modulation phase, we can control the phase difference between the incident and reflected waves. Such reflection is equivalent to the reflection from a static boundary modeled by a parallel connection of resistive and reactive elements. Therefore, the response described by the complex reflection coefficient at each input frequency can be interpreted by the static equivalent circuit shown in Fig.~\ref{fig1c}. Doing some algebraic manipulations, the effective parameters of this equivalent circuit at each of the input frequencies $\omega_i$ can be written as
\begin{equation}
\begin{split}
&G_{\omega_i}^{\rm eq}=\frac{2\left(\Gamma_{\omega_i}^{\rm (R)}+1\right)}{\eta_0\left(\Gamma_{\omega_i}^{\rm (R)}+1\right)^2+\eta_0 {\Gamma_{\omega_i}^{\rm (I)}}^2} -\frac{1}{\eta_0},\\
&B_{\omega_i}^{\rm eq}=-\frac{2\Gamma_{\omega_i}^{\rm (I)}}{\eta_0\left(\Gamma_{\omega_i}^{\rm (R)}+1\right)^2+\eta_0 {\Gamma_{\omega_i}^{\rm (I)}}^2},
\end{split}
\end{equation}
where $\Gamma_{\omega_i}^{\rm (R)}$ and $\Gamma_{\omega_i}^{\rm (I)}$ are the real and imaginary parts of the reflection coefficient at the corresponding input frequency. We see that  modulation of a resistive boundary creates a virtual reactive component that can be fully controlled by engineering the modulation parameters. Notice that the virtual reactive component should not be interpreted as a capacitive or inductive load. It is important to stress that there is no energy storage in the system, and the existence of a virtual reactive component in the equivalent circuit has its origin from the phase delay created as a consequence of the interference between the input harmonics and the modulation products. We see that the effective parameters depend on the reflection coefficient at the same frequency, and, interestingly, they can have different values at different frequencies as the reflection can be asymmetric. Consequently, designing the modulation parameters gives us an opportunity to fully control reflection as if there were different circuit elements for input signals at different frequencies. This is an extremely powerful functionality, as it can be used to match a boundary at multiple frequencies simultaneously. 

\begin{figure}[t!]
\centering
\subfigure[]{\includegraphics[width=0.5\linewidth]{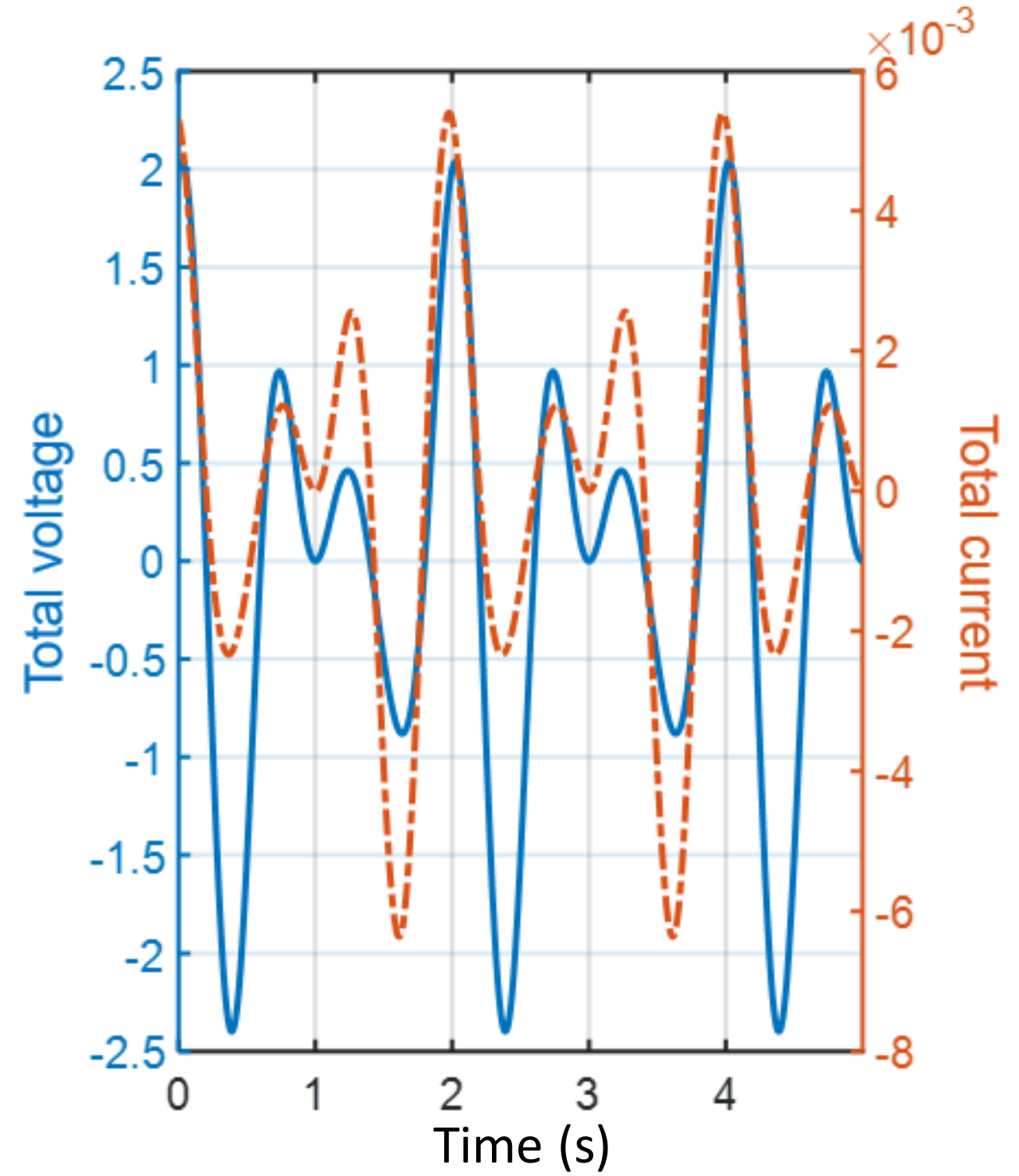} \label{fig5a}}%
\subfigure[]{\includegraphics[width=0.48\linewidth]{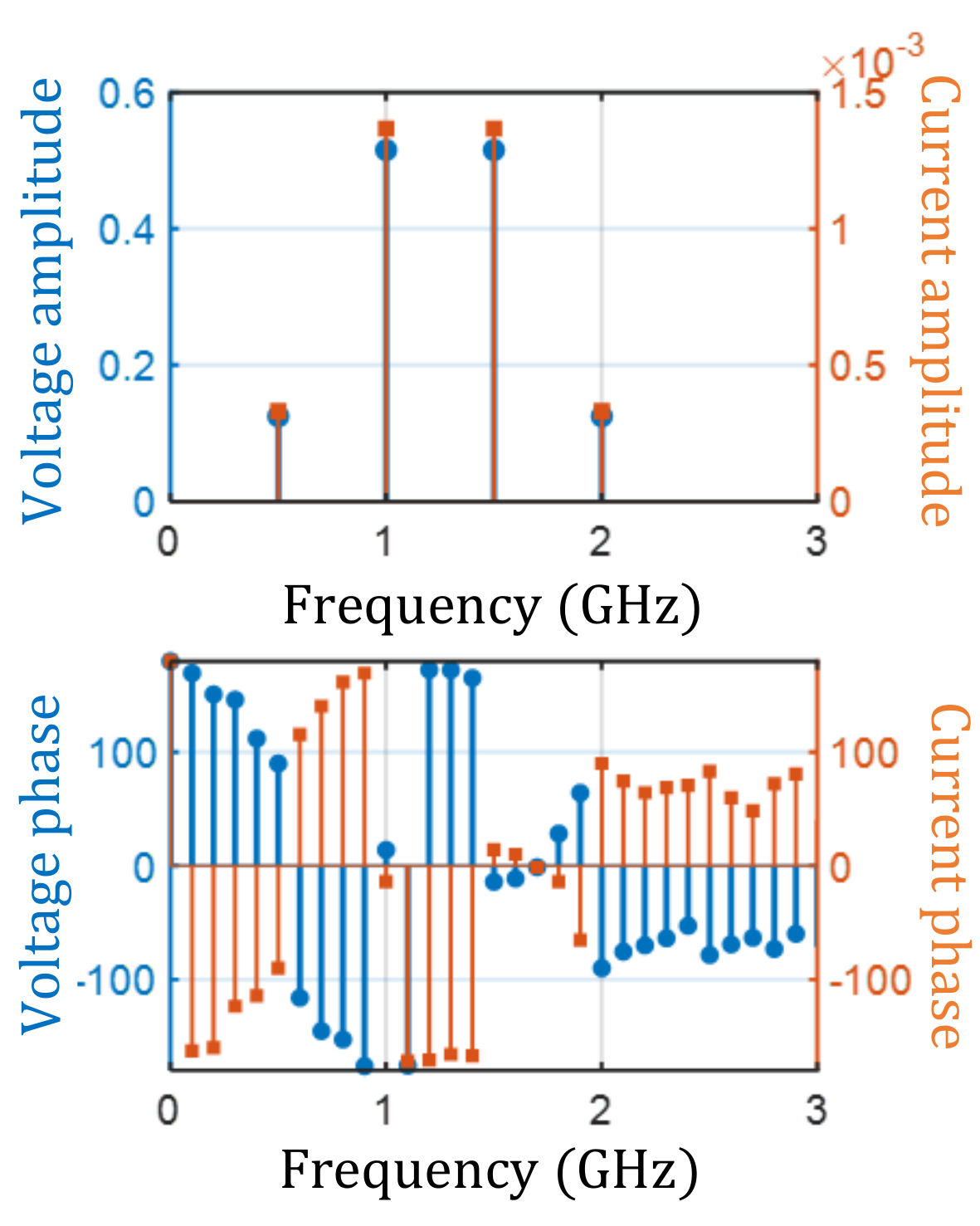}\label{fig5b}}%
\caption{Current distortion due to modulation. 
(a) Temporal variation of the total voltages and currents. (b) Amplitude and phase of the total currents and voltages in frequency domain.} \label{fig:Fig5}
\end{figure}

For better understanding of these phenomena, we can analyze the total voltages and currents that are represented in Fig.~\ref{fig:Fig5} when the modulation function is defined by $f_{\rm m}=0.5$~GHz, $\chi_{\rm 0}=1$, $\chi_{\rm m}=0.5$, and $\phi_{\rm m}=\pi/2$, and the input signals by $f_1=1$~GHz, $f_2=1.5$~GHz, $a_1=a_2=1$~V, and $\phi_1=\phi_2=0$. Figure~\ref{fig5a} shows a comparison of the temporal variation of the total voltage and current where we can see the effect produced by the harmonic distortion. For a deeper study of the harmonics, we represent the total voltages and currents in the frequency domain, see Fig.~\ref{fig5b}. We observe phase differences between the voltages and currents of each harmonic that are responsible for current distortions. To evaluate the effect of the harmonic distortion on the total power dissipated in the boundary, we calculate the power carried by the reflected harmonics. The individual average power transported by each harmonic can be written as
\begin{equation}
\begin{split}
&P^-_{\omega_1}=\frac{a_1^2}{2\eta_0}\left[(1-\chi_0)^2+\frac{\chi_{\rm m}^2}{4} \frac{a_2^2}{a_1^2}-\frac{a_2}{a_1} (1-\chi_0)\chi_{\rm m}\cos{\Phi}\right],\cr
&P^-_{\omega_2}=\frac{a_2^2}{2\eta_0}\left[(1-\chi_0)^2+\frac{\chi_{\rm m}^2}{4} \frac{a_1^2}{a_2^2}- \frac{a_1}{a_2} (1-\chi_0)\chi_{\rm m}\cos{\Phi}\right],\cr
&P^-_{\omega_2-2\omega_1}= \frac{a_1^2}{2\eta_0}\frac{\chi_{\rm m}^2}{4},\cr
&P^-_{2\omega_2-\omega_1}= \frac{a_2^2}{2\eta_0}\frac{\chi_{\rm m}^2}{4},
\label{POWERADH}
\end{split}
\end{equation}
where $\Phi=\phi_1+\phi_{\rm m}-\phi_2$. From this equation, we see how the reflected power, and, consequently, the absorption in the resistive layer are modified by the time modulation. The final expression for the absorbed power $A$ is found by subtracting the total average power carried out by the harmonics of the reflected waves from the total average incident power. This way we find that
\begin{equation}
A={\left(2\chi_0-\chi_0^2-{\chi_{\rm m}^2\over2}\right)(a_1^2+a_2^2)+2a_1a_2(1-\chi_0)\chi_{\rm m}\cos{\Phi}\over2\eta_0}.
\end{equation}
Absorption phenomena in this system are similar to what happens in networks with nonlinear loads where the generation of spurious harmonics (e.g., $P^-_{\omega_2-2\omega_1}$ and $P^-_{2\omega_2-\omega_1}$) reduces the power factor, and, consequently, reduces the active power absorbed by the loads. Thus, to assure the maximum absorption, spurious reflected harmonics should be minimized, thus, $\chi_{\rm m}$ has to be relatively small.

It is also important to note that the modulated resistive boundary is lossy as long as the conditions in Eq.~\eqref{conditions} are met. These conditions assure that the resistance of the boundary is always positive, meaning that at every moment of time Ohm's law reads $i_{\rm tot}(t)=v_{\rm tot}(t)/r(t)$ with the voltage and current having the same sign, which implies instantaneous power dissipation rather than power generation. A time-modulated positive resistance supports only forced oscillations, and it is not possible to sustain free oscillations in a circuit with total positive resistance. This is confirmed by Eq.~\eqref{eq:reflectionco}, 
where the instantaneous reflection coefficient can have values $-1<\gamma(t)<1$ for positive values of $g(t)$. It can have values outside this range only when the resistive boundary has a negative value, in which case the resistive boundary is introducing gain to the incident signal. 

For time-varying reactive elements, there is one term added in the voltage-current relation due to the non-zero capacitance/inductance time derivative, and this term indicates power gain/loss due to the exchange of power between the main circuit and the modulation circuit. However, for time-varying positive resistive elements, Ohm's law has the same formulation as for static resistors, indicating that it is possible to modulate resistive elements without exchanging power with the modulation circuit (see Appendix~\ref{appd}).
	

\section{Multi-frequency perfect absorption by ultra-thin metasurfaces}
\label{abcs}

\begin{figure*}[]
\centering
 \begin{minipage}[]{0.28\textwidth}
\subfigure[]{\includegraphics[width=1\linewidth]{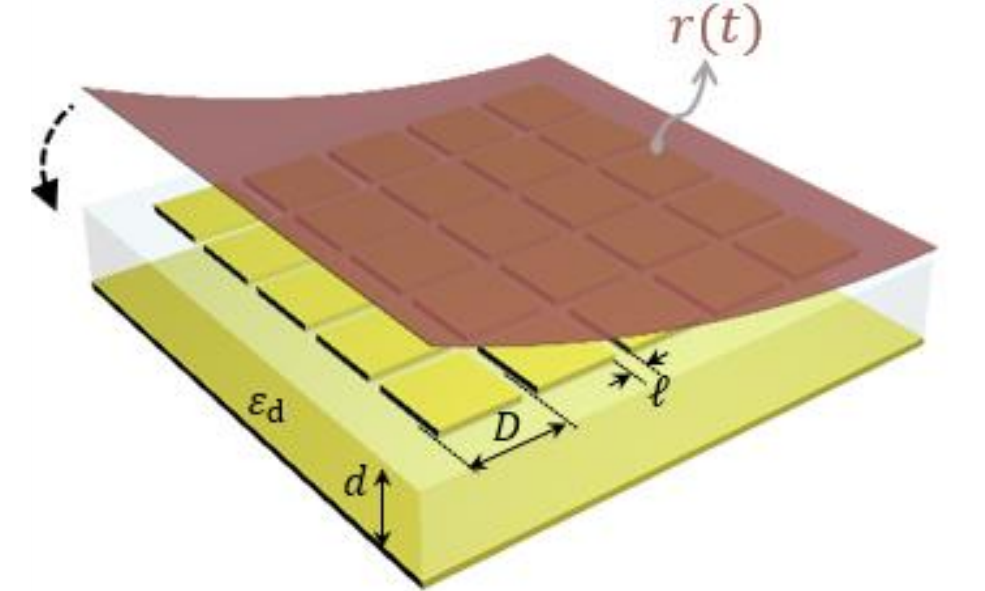} \label{fig3a}}
\end{minipage}
\begin{minipage}[]{0.33\textwidth}
\subfigure[]{\includegraphics[width=0.45\linewidth]{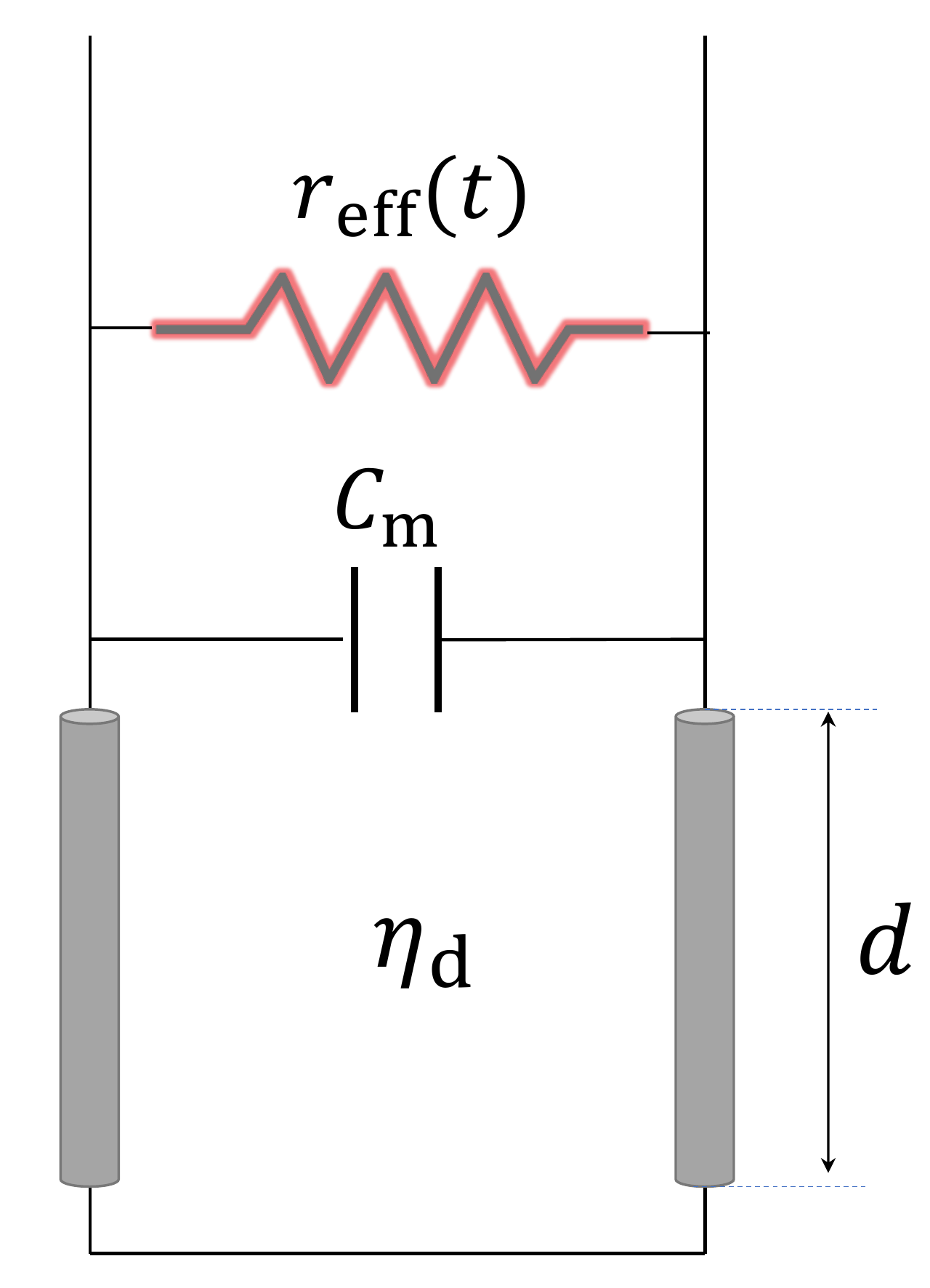}\label{fig3b}}
\subfigure[]{\includegraphics[width=0.47\linewidth]{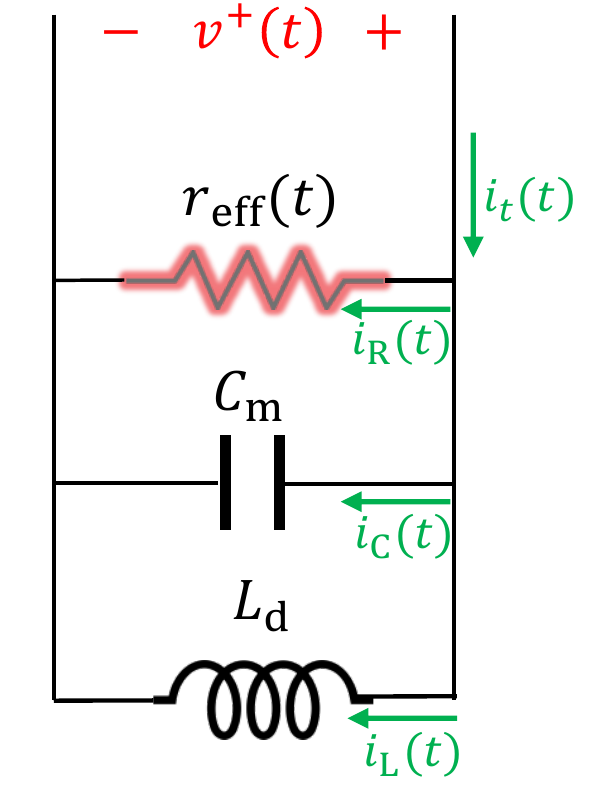}\label{fig3c}}
\end{minipage}
\begin{minipage}[]{0.33\textwidth}
 \subfigure[]{\includegraphics[width=1.2\linewidth]{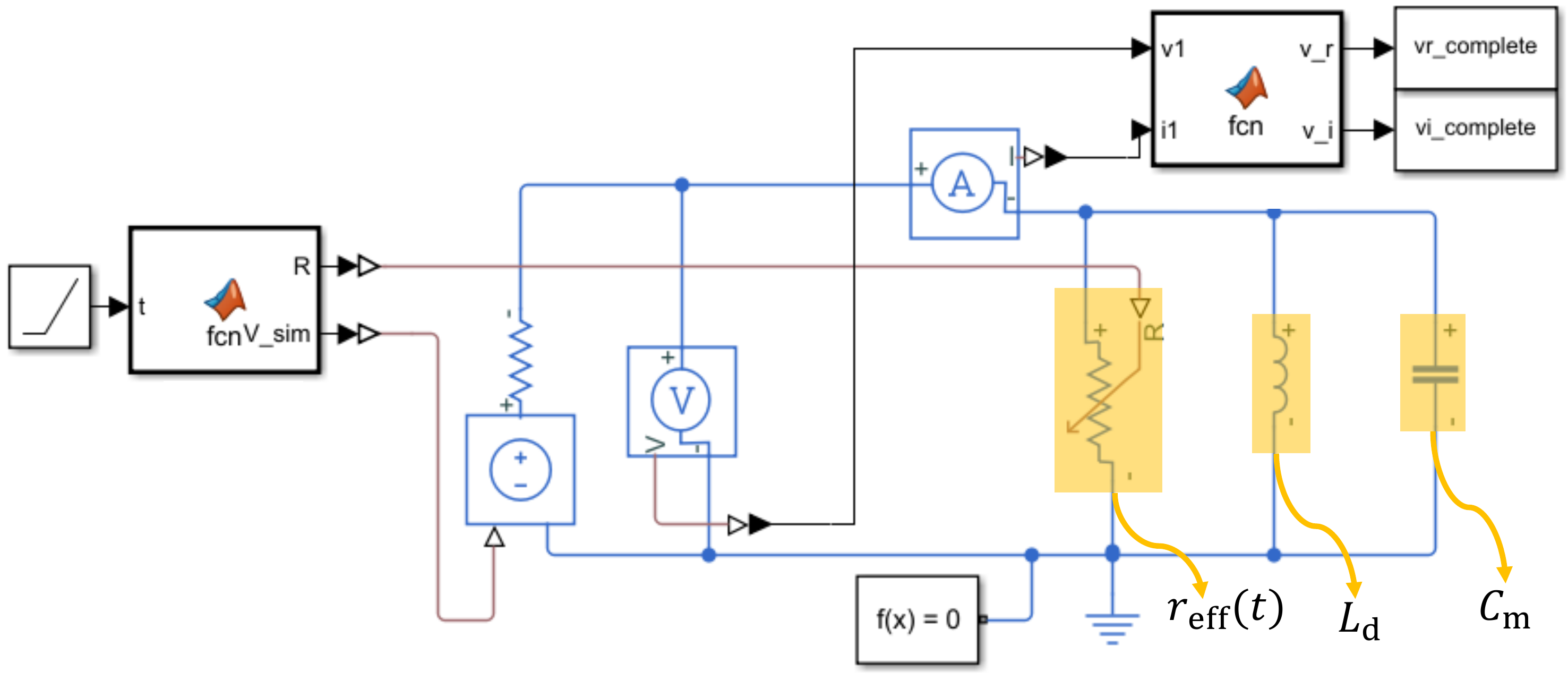} \label{simulink}}
\end{minipage}
\begin{minipage}[]{0.3\textwidth}
\subfigure[]{\includegraphics[width=1\linewidth]{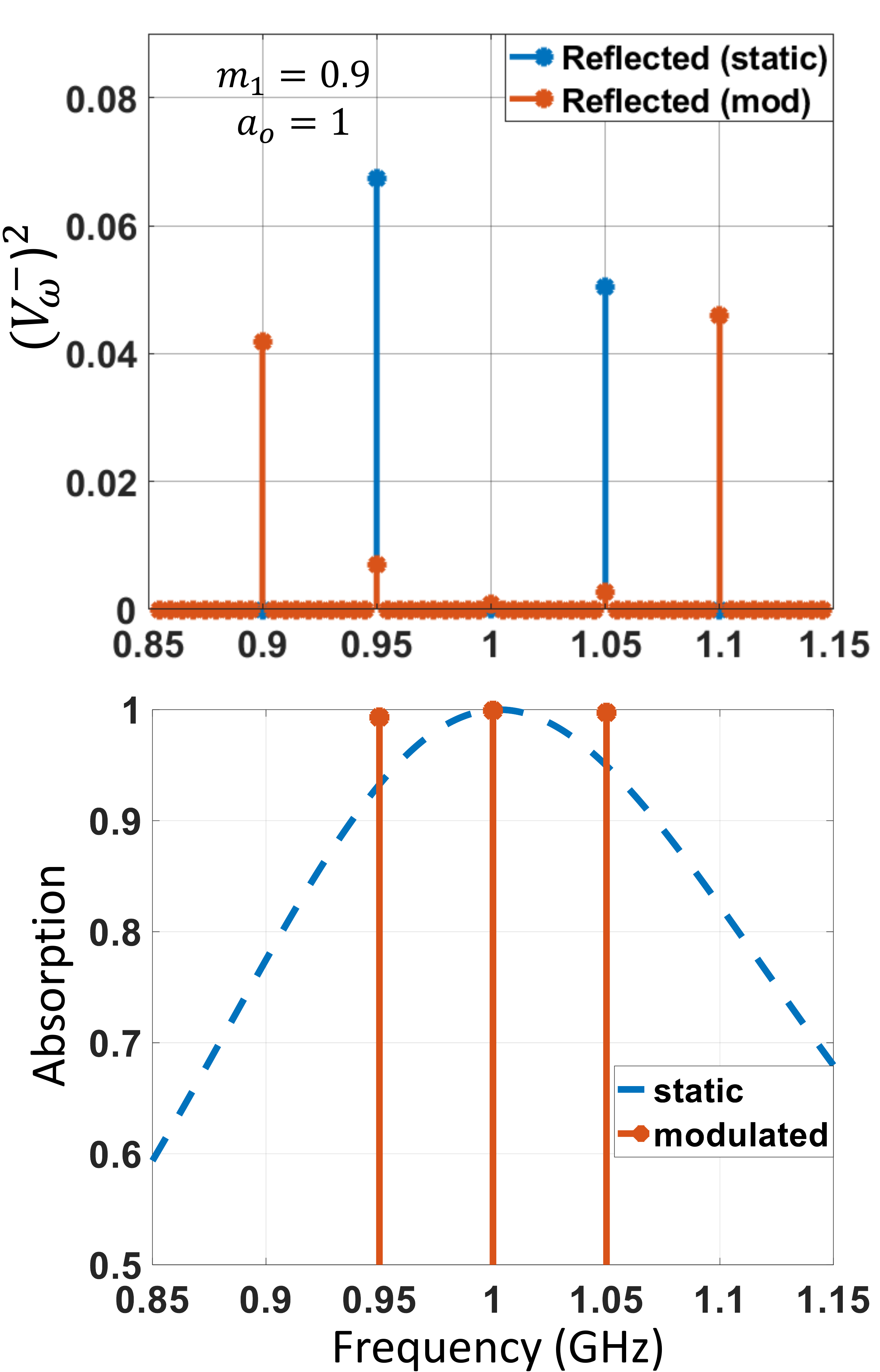} \label{fig3d}}
\end{minipage}
\begin{minipage}[]{0.3\textwidth}
\subfigure[]{\includegraphics[width=1\linewidth]{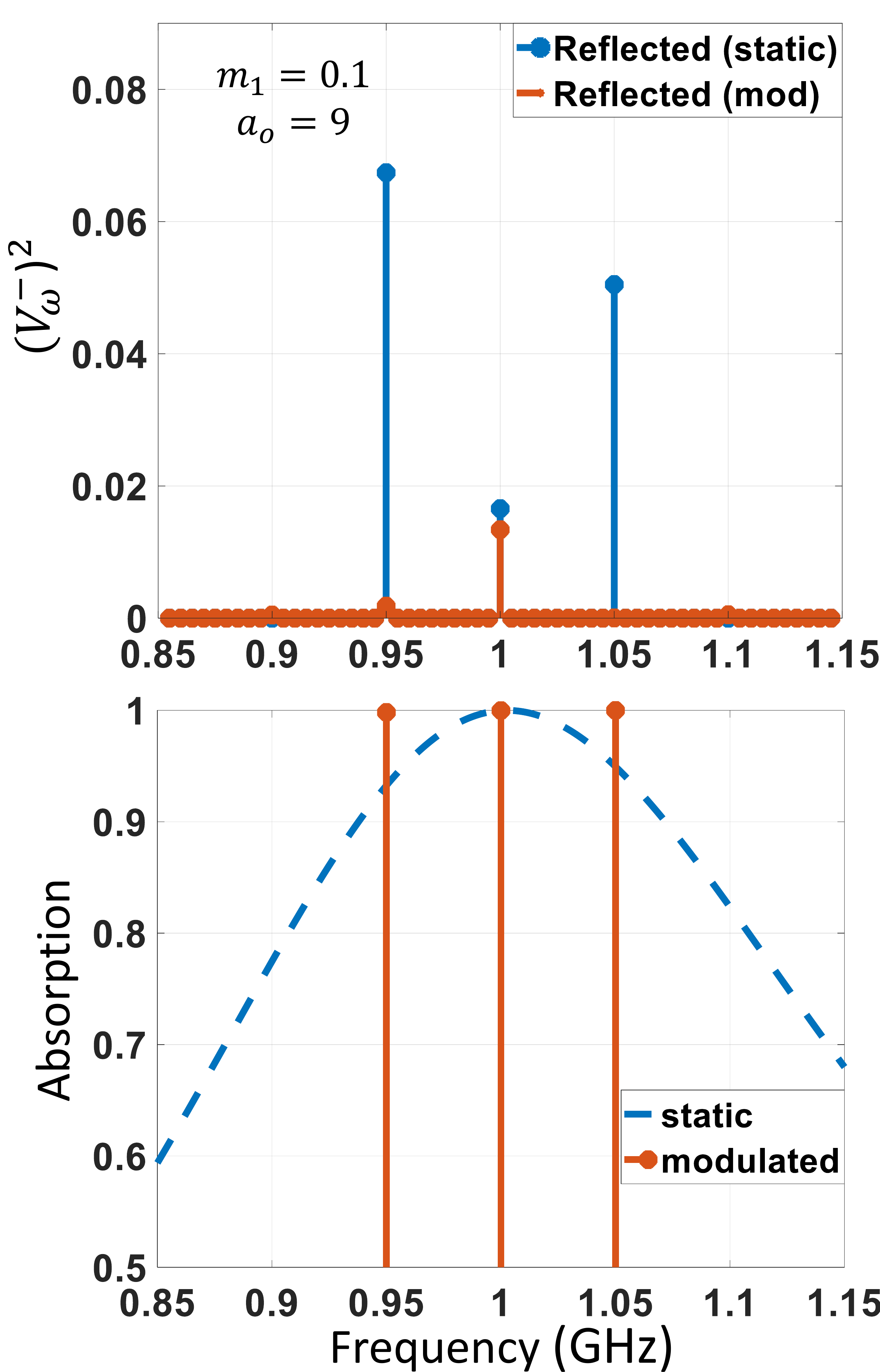} \label{fig3e}}
\end{minipage}
\caption{Impenetrable metasurface with a time-modulated resistive layer. (a) Schematic representation of a time-varying reflective metasurface. (b) Equivalent circuit model of the proposed structure for
normal incidence. (c) Simplified version of the equivalent model and current distribution. (d) Simulation schematic from SIMULINK. (e) Comparison between the reflected voltage’s amplitude squared as a function of frequency in the static and dynamic cases when $m_1=0.9$ and $a_0=1$~V. (f) Comparison between the reflected voltage’s amplitude squared as a function of frequency in the static and dynamic cases when $m_1=0.1$ and $a_0=9$~V.} 
\label{cap}
\end{figure*}

To understand the possibilities offered by multi-frequency illumination of  coherently time-modulated surfaces, let us consider a realistic metasurface structure (see Appendix~\ref{appc}) shown in Fig.~\ref{cap}(a). This reflective metasurface consists of
a grounded dielectric layer with the relative permittivity $\varepsilon_{\rm d}$ and thickness $d$. A periodical array of thin metal patches (with an electrically small period) is positioned on top of the substrate, and, over the patches, a dispersionless time-varying resistive sheet is placed. A dispersionless thin resistive sheet for microwave applications can be realized as a thin conductive (metal) sheet, see e.g.~\cite{sergeiblue}. The sheet resistance of  thin layers of conductors is $\frac{1}{\sigma d_{\rm{r}}}$, where $\sigma$ is the material conductivity and $d_{\rm{r}}$ is the sheet thickness. Since conductivity of metals is constant in wide frequency ranges, the sheet resistance does not depend on frequency, and it can be modulated by tuning the conductivity (see Appendix~\ref{appc}). Such metasurface can be modeled by the equivalent circuit in Fig.~\ref{cap}(b). The grounded substrate is represented by a shorted transmission line with the characteristic impedance $\eta_{\rm d}=\sqrt{\mu_0/(\varepsilon_0\varepsilon_{\rm d}})$. Considering normal incidence, the equivalent shorted transmission line can be viewed as a reactive load whose impedance equals $Z_{\rm d} =j\eta_{\rm d} \tan(k_{\rm d}d)$, where $k_{\rm d} = \omega \sqrt{\mu_0\varepsilon_0\varepsilon_{\rm d}}$. 
The patch array has a capacitive behaviour, and it can be modeled by the sheet capacitance $C_{\rm m}=2\epsilon_{\rm eff}\varepsilon_0D \ln{\left(\csc\frac{\pi}{2(p+1)}\right)}/\pi$, where $\epsilon_{\rm eff}=(\epsilon_{\rm d}+1)/2$ and $p=(D-\ell)/\ell$.
Finally, with the presence of the patch array, the resistive layer is partially shorted and has an effective sheet resistance $r_{\rm eff}(t)=r(t)\ell/(D-\ell)$ \cite{wang2018toward}.

Next, we analyse this metasurface when it is illuminated by a multifrequency periodical input signal defined by
$v^+(t)=\sum_{n=-N}^{N} a_n \cos(\omega_{n}t)$, where $2N+1$ is the number of input harmonics. We consider spectra formed by several harmonics symmetrically located at the two sides of the resonance frequency of the structure that satisfies $\omega_0=(\omega_n+\omega_{-n})/2$. The harmonic at $\omega_0$ we call \emph{control signal}, because it can be used to control absorption at other frequencies. In this paper, we analyze the response of the structure when the effective conductance of the lossy layer varies according to $g_{\rm eff}(t)=1/r_{\rm eff}(t)=g_0+\sum_{k=1}^{N} g_0 m_k \cos(\omega_{{\rm m}k}t+\phi_{{\rm m}k})$, where $m_k$, $\phi_{{\rm m}k}$ are the modulation depths and phases, and $\omega_{{\rm m}|n|}=\omega_{|n|}-\omega_{0}$ are the frequencies of the modulation harmonics. We will show that by properly tuning the modulation parameters and the control signal's amplitude $a_0$, perfect absorption at all input frequencies can be achieved even for ultrathin metasurfaces. 

In the previous section, we analysed how the modulation of the resistive boundary modifies the reflection coefficient, which was a powerful formulation that allowed us to properly analyse the scattering properties of the structure. However, the structure that we study in this section, is also reactive, and analytically formulating the reflection coefficient for a boundary consisting of a time-varying resistive layer in parallel with reactive layers is mathematically complicated. To keep the mathematical formulation as simple as possible, we perform the analysis differently. Let us assume that the structure is reflectionless, that is, the voltage at the input port equals to the incident voltage, as shown in Fig.~\ref{fig3c}. In this case, the current in the resistive sheet can be written in terms of the time-varying effective conductance as 
\begin{equation}
\begin{split}
i_{\rm R}(t)=&\overbrace{\sum_{n=-N}^{N} a_n g_0 \cos(\omega_{n}t)}^{T_1}\cr
&+\overbrace{\sum_{n=-N}^{N} a_n \cos(\omega_{n}t) \sum_{k=1}^{N} g_0 m_k \cos(\omega_{{\rm m}k}t+\phi_{{\rm m}k})}^{T_2}.
\end{split}
\end{equation}
Here, $T_1$ is the current produced in the resistive element in case of no modulation applied, and $T_2$ is the current produced due to modulation (modulation products).

To analyse the scattering properties, we need to calculate currents flowing in the reactive layers. If we consider the grounded substrate to be electrically thin, we can use the approximation $Z_{\rm d}\approx j\omega L_{\rm d}$ where $L_{\rm d}=\mu_0d$. Hence, the total electric current through this effective inductance and the capacitive sheet modeling the metallic patches is written as
\begin{equation}
i_{\rm L}(t)+i_{\rm C}(t)= \sum_{n=-N}^{N} \left[\frac{1}{L_{\rm d} \omega_{n}}- C_{\rm m} \omega_{n}\right] a_n \sin(\omega_{n}t),
\end{equation}
where $\frac{-1}{L_{\rm d} \omega_{n}}$ and $C_{\rm m} \omega_{n}$ are the corresponding susceptances for the inductance and capacitance, respectively, and $B_n=\frac{-1}{L_{\rm d} \omega_{n}}+ C_{\rm m} \omega_{n}$ is the total susceptance. As for any parallel resonator, at frequencies $\omega_k$, the total susceptances $B_k$ are positive (capacitive), and at frequencies $\omega_{-k}$, the total susceptances $B_{-k}$ are negative (inductive). For simplicity, we assume that $|B_k|=|B_{-k}|$, therefore, they are equal in magnitude but have the opposite signs. Finally, according to Fig.~\ref{fig3c}, the total electric current reads $i_{t}(t)=T_1+T_2+i_{\rm L}(t)+i_{\rm C}(t)$. Assuming that $g_0=1/\eta_0$, and spurious harmonics (defined as the frequency components not present in the incident signal, but excited due to modulation) are negligible, the condition for perfect total absorption is met when $i_{\rm L}(t)+i_{\rm C}(t)=-T_2$. The frequency mixing that takes place in $T_2$ produces modulation products with frequencies $\omega_{\pm k}$. To obtain perfect absorption, every modulation product should compensate the reactive current at the corresponding frequency. As a result, the condition for perfect absorption can be reformulated in the frequency domain as $I_{\rm L}(\omega_n)+I_{\rm C}(\omega_n)=-T_2(\omega_n)$, that is satisfied when
\begin{equation}
\phi_{{\rm m}|n|}=\frac{3 \pi}{2},\,\,\,\,\,\,\,\,\frac{\vert a_{0} g_0 m_{|n|}\vert}{2} = \left|\Big[\frac{1}{L_{\rm d} \omega_{n}}- C_{\rm m} \omega_{n}\Big] a_{n}\right|.
\label{eq_coditions}
\end{equation}
As the reactive currents $i_{\rm L}(t)+i_{\rm C}(t)$ are antisymmetric around the resonance frequency, having negative values at frequencies $\omega_k$ and positive values at frequencies $\omega_{-k}$, the generated modulation products should be also antisymmetric (with the opposite sign convention). The required antisymmetry is obtained by choosing the modulation phase $\phi_{{\rm m}|n|}=\frac{3 \pi}{2}$. As a result, modulation products at frequencies $\omega_{k}$ are positive, and modulation products at frequencies $\omega_{-k}$ are negative. Thus, modulation products are equal in magnitude with the reactive currents and have the opposite sign, so they cancel each other, and perfect absorption at multiple frequencies is achieved. In addition, due to the antisymmetry, the modulation products produced at the resonance frequency $\omega_0$ are equal and have the opposite signs, so they cancel out. Thus, the signal at the resonance frequency is not affected by modulation. Notice that the preceding calculations are accurate only when the amplitudes of the spurious harmonics are negligibly small. Otherwise, it will be impossible to satisfy $i_{\rm L}(t)+i_{\rm C}(t)=-T_2$, as $T_2$ has spurious frequency components that have no equivalency in $i_{\rm L}(t)+i_{\rm C}(t)$. To keep spurious harmonics negligibly small, $m_k$ should be relatively small. This condition can be ensured without compromising performance, as $a_0$ can be increased to obtain the desired performance while maintaining $m_k$ small.

Importantly, these results are obtained using slow modulation, where the modulation frequency is much smaller than all the input frequencies, which is more practical than the double frequency modulation usually used in parametric devices. In addition, it does not matter how large the reactive currents are, we can always obtain perfect absorption. This can be seen from Eq.~(\ref{eq_coditions}) where the modulation products' magnitudes depend on $a_0$, which has no theoretical limit and can be as large as desired. As a result, this technique works for ultrathin metasurfaces: no matter how large is the difference $\omega_n-\omega_0$, perfect absorption can be obtained at $\omega_n$.


To validate the design conditions obtained above, we target a metasurface to absorb incident power at three frequencies: 0.95~GHz, 1~GHz, and 1.05~GHz, where 1~GHz is the control frequency. In this scenario, there is only one modulation term with the frequency $\omega_{m1}=2 \pi 50 \times 10^6 $~rad/sec, modulation depth $m_1$, and modulation phase $\phi_{m1}$. Next, we select easily realizable structure parameters: the permittivity of the grounded dielectric layer is $\varepsilon_d=12$, and the thickness $d=9.54 \: {\rm mm}\approx\lambda/31$ (the conditions that are equivalent to an inductive response defined by $L_{\rm d}= 12 \: {\rm nH}$). An array of metal patches is designed to produce a resonance at the control frequency, which corresponds to the period of $D=\lambda_0/10=30$~mm and $d=2.83 \:{\rm mm}$ (the equivalent capacitance $C_{\rm m}=2.1 \: {\rm pF}$). According to the condition for having perfect absorption at all three frequencies [see Eq.~(\ref{eq_coditions})], the amplitude of the control signal and the modulation depth must satisfy $a_{0} m_1 \approx 0.9 a_{\pm 1}$. To verify the design, we first assume that all the input harmonics have the same amplitude $a_0=a_{\pm 1}$, and the amplitude and phase of the modulation function are $m_1=0.9$ and $\phi_{m1}=\frac{3 \pi}{2}$, respectively, which satisfies the condition in Eq.~(\ref{eq_coditions}). Next, we simulate the equivalent circuit presented in Fig.~\ref{fig3c} using SIMULINK,  as presented in Fig.~\ref{simulink}. The incident and reflected voltage signals are extracted from SIMULINK, then the absorption and the reflected voltage's amplitude squared as a function of frequency are calculated, and the results are shown in Fig.~\ref{fig3d}. We can see that the reflected power at the input frequencies is reduced by the modulation. However, because the modulation depth $m_1$ is not small, the power of the spurious harmonics is not negligible, as expected from the initial analysis.

\begin{figure*}[]
\centering
\subfigure[]{\includegraphics[width=0.42\linewidth]{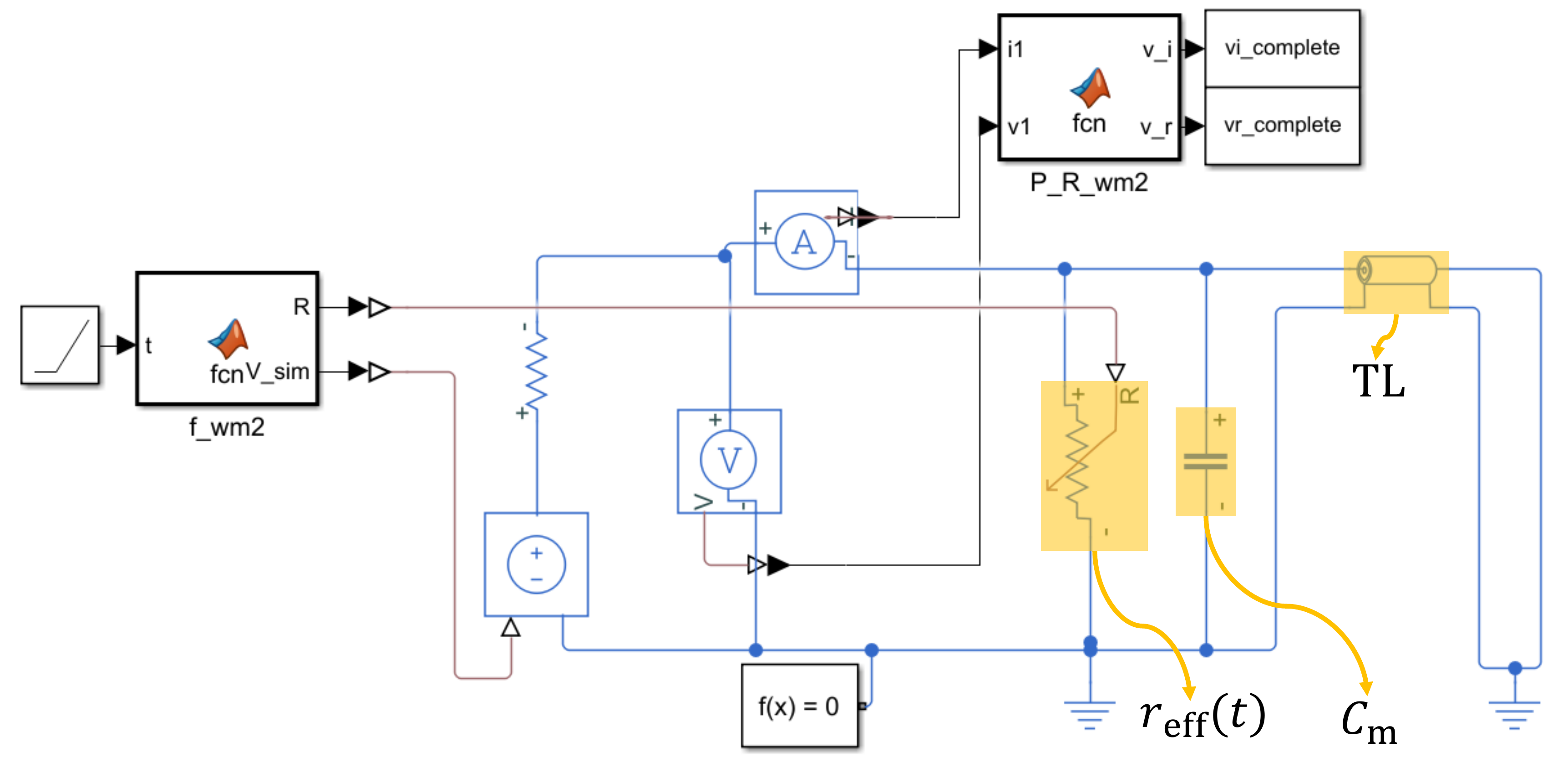}\label{fig:sim2}}
\hspace{2em}%
\subfigure[]{\includegraphics[width=0.30\linewidth]{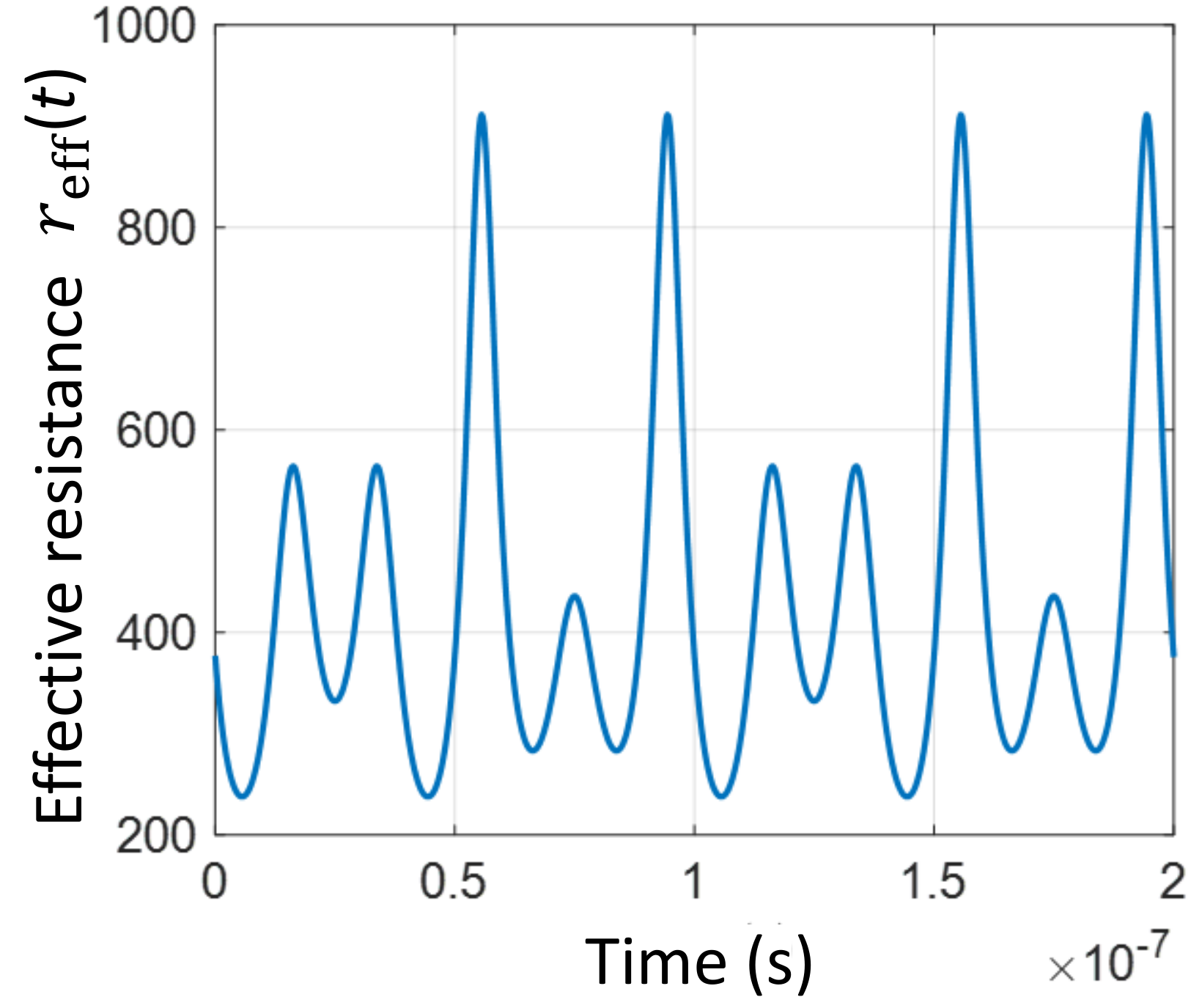}\label{fig:fig6a}}
\hspace{2em}%
\subfigure[]{\includegraphics[width=0.34\linewidth]{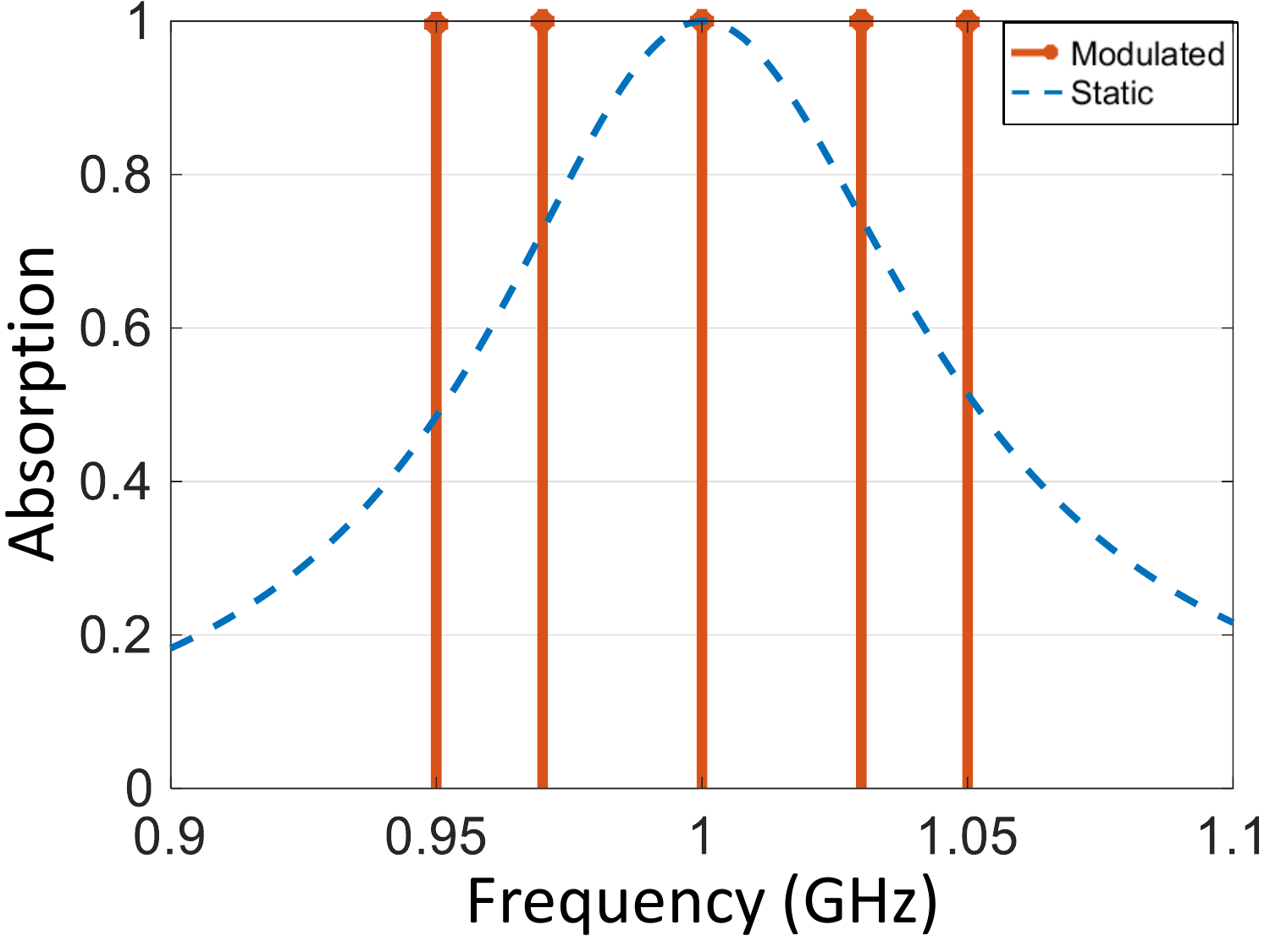}\label{fig:fig6b}}
\hspace{2em}%
\subfigure[]{\includegraphics[width=0.35\linewidth]{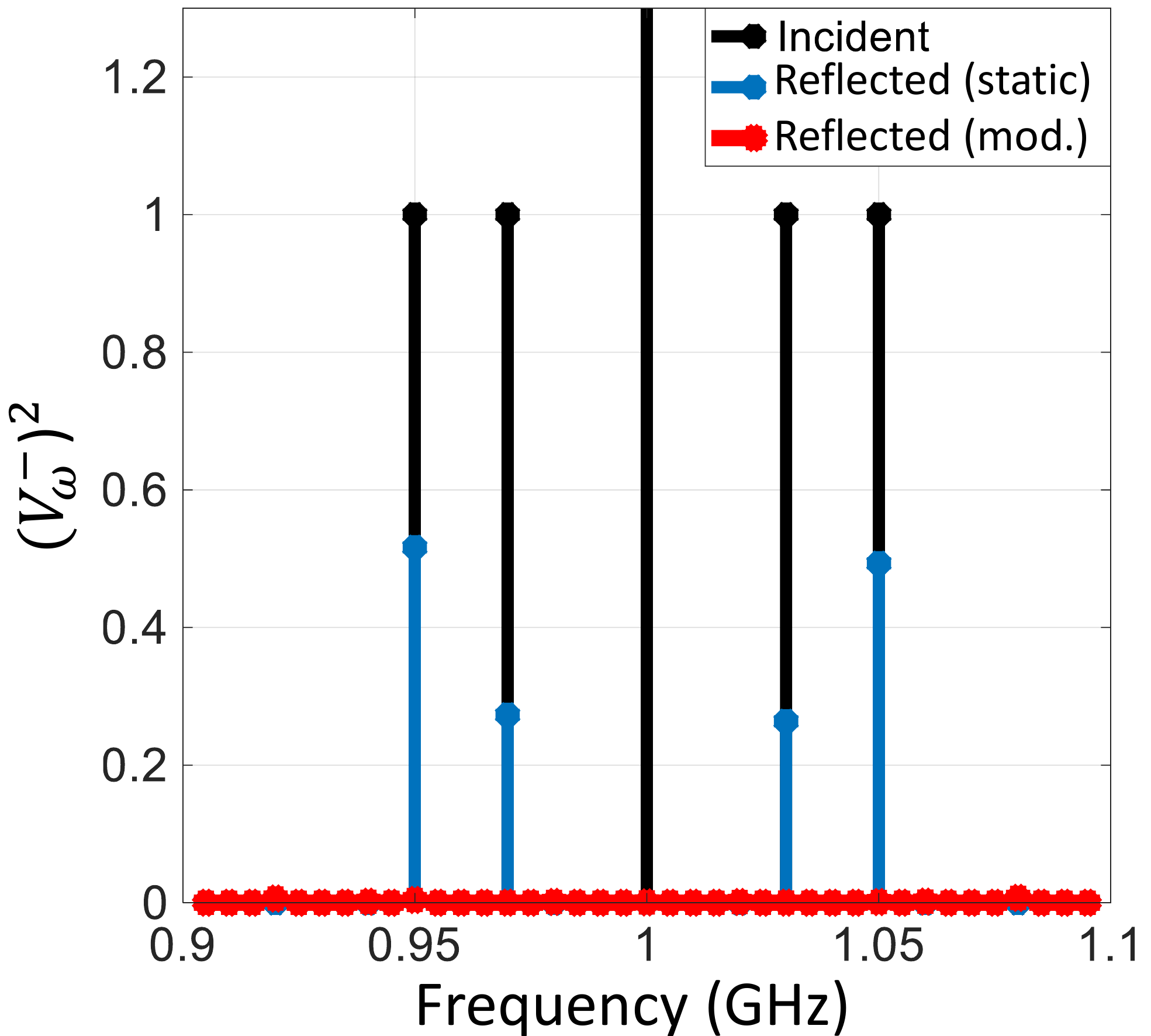}\label{fig:fig6d}}
\hspace{2em}%
\subfigure[]{\includegraphics[width=0.36\linewidth]{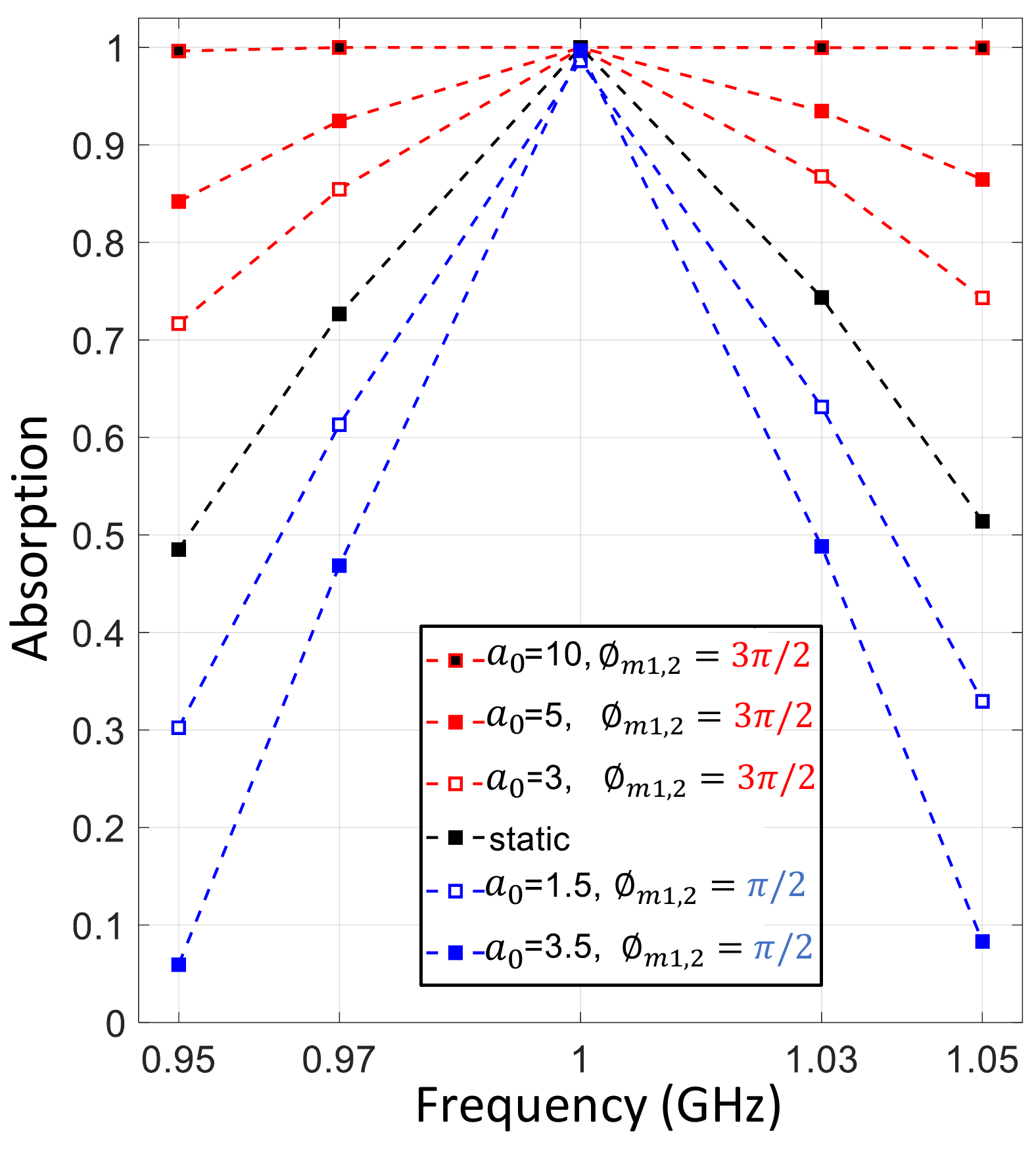}\label{fig:fig6e}}
\hspace{2em}%
\subfigure[]{\includegraphics[width=0.33\linewidth]{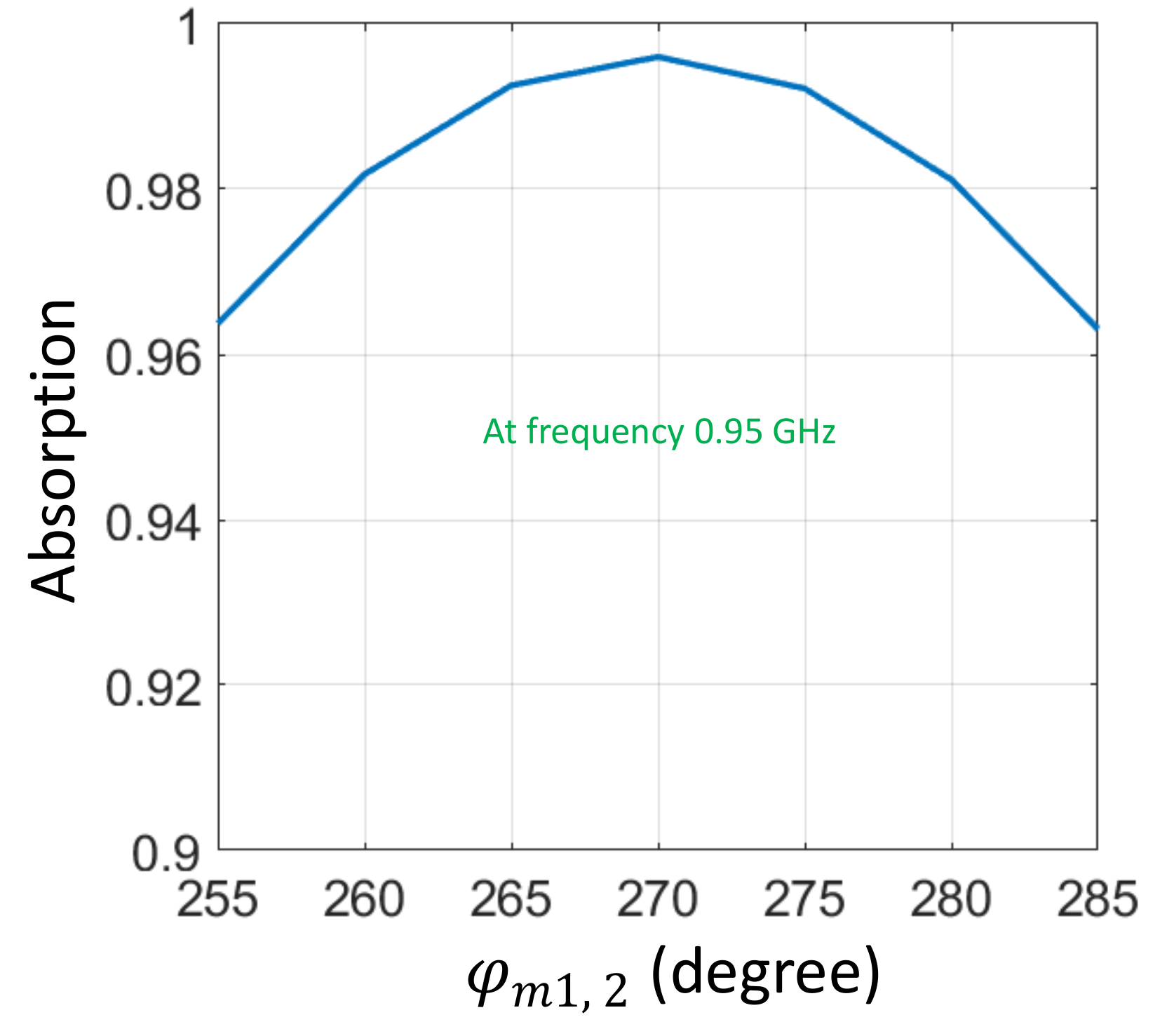}\label{fig:fig6c}}
\caption{Perfect absorption, remote tunability, and robustness in dynamic metasurfaces. (a) SIMULINK simulation schematic. In this schematic a transmission line with $\varepsilon_d=12$ and $d=2.367$~mm is used to model the dielectric layer. (b) The effective resistance required to obtain perfect absorption at 0.95 GHz, 0.97 GHz, 1 GHz, 1.03 GHz, and 1.05 GHz when $a_{0}=10$~V and $a_{\pm2}=a_{\pm 1}=1$~V. (c) Comparison between the absorption of the static and dynamic structures.  (d) Comparison between the reflected voltage amplitude squared as a function of the frequency for static and dynamic structures.
(e) Tunability of absorption by controlling the amplitude of the control signal and the modulation phase. (f) Absorption at frequency 0.95~GHz as a function of the modulation phase.}
 \label{fig_Fig6}
\end{figure*}

To reduce the power of the spurious harmonics in the reflected spectrum, we can consider an alternative scenario where $a_0=9a_{\pm 1}$ and $m_1=0.1$. The reflected voltage amplitude squared as a function of frequency for this configuration is presented in Fig.~\ref{fig3e}, where we can see that in this case the power of all spurious harmonics is drastically reduced, and the structure behaves as a perfect absorber at all input frequencies. If we compare the absorption of the input harmonics with the equivalent static absorber, we see a considerable improvement. This example shows the significance of the multi-frequency illumination, as controlling $a_0$ allows us to boost the performance. Such boost cannot be obtained in conventional parametric devices, as the performance depends only on the modulation depth, which is limited in most cases. It is important to stress that the benefits of this technique become more significant for thin metasurface absorbers, where the resonant system created by the grounded dielectric and the array of patches produces a narrow-band resonance. As it is described by Eq.~(\ref{eq_coditions}), for thinner absorbers with large reactive currents, the amplitude of the control signal $a_0$ can be increased as much as desired to obtain perfect absorption. 

\begin{figure*}[]
\centering
\subfigure[]{\includegraphics[width=0.28\linewidth]{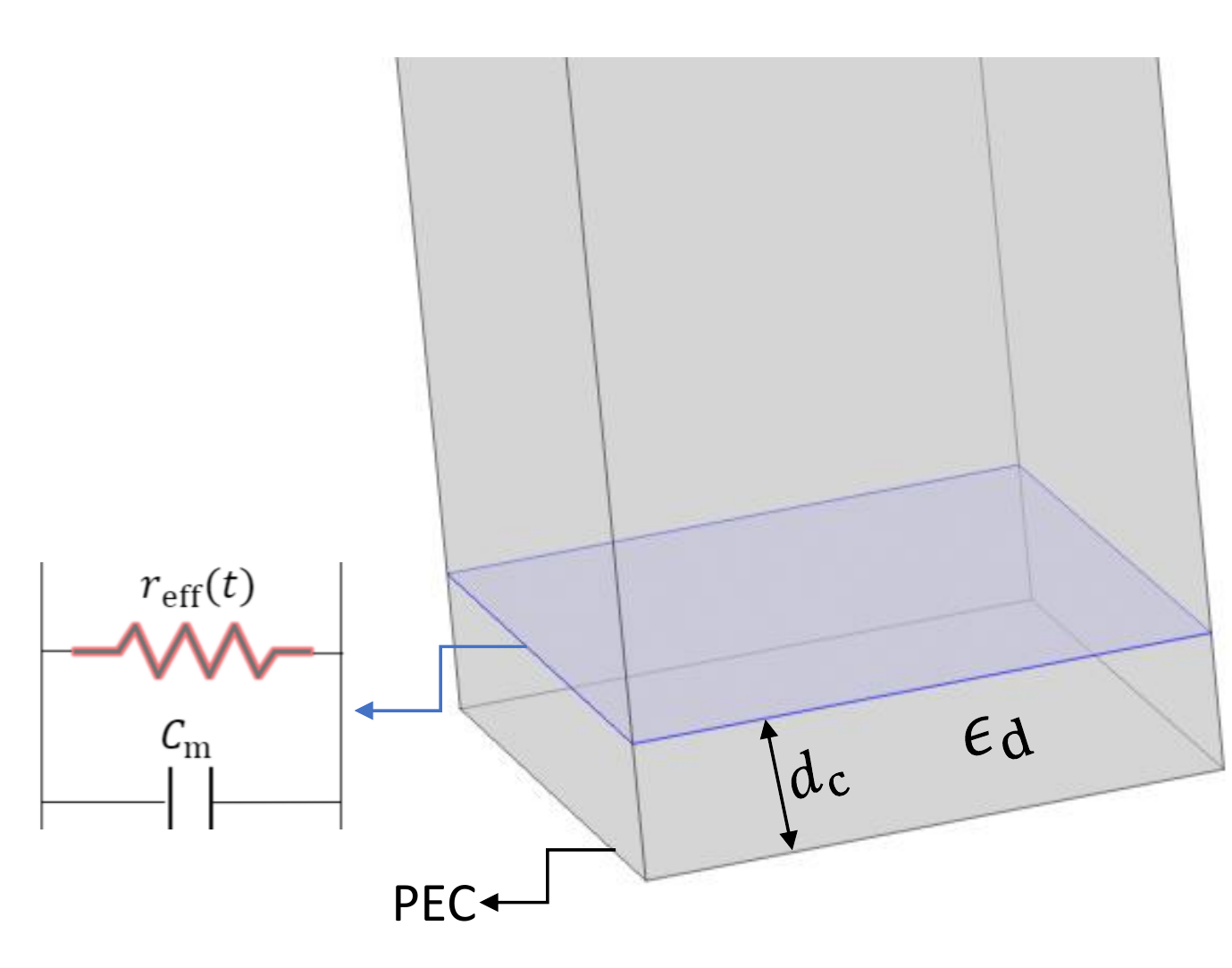}\label{fig:fig7a}}
\subfigure[]{\includegraphics[width=0.33\linewidth]{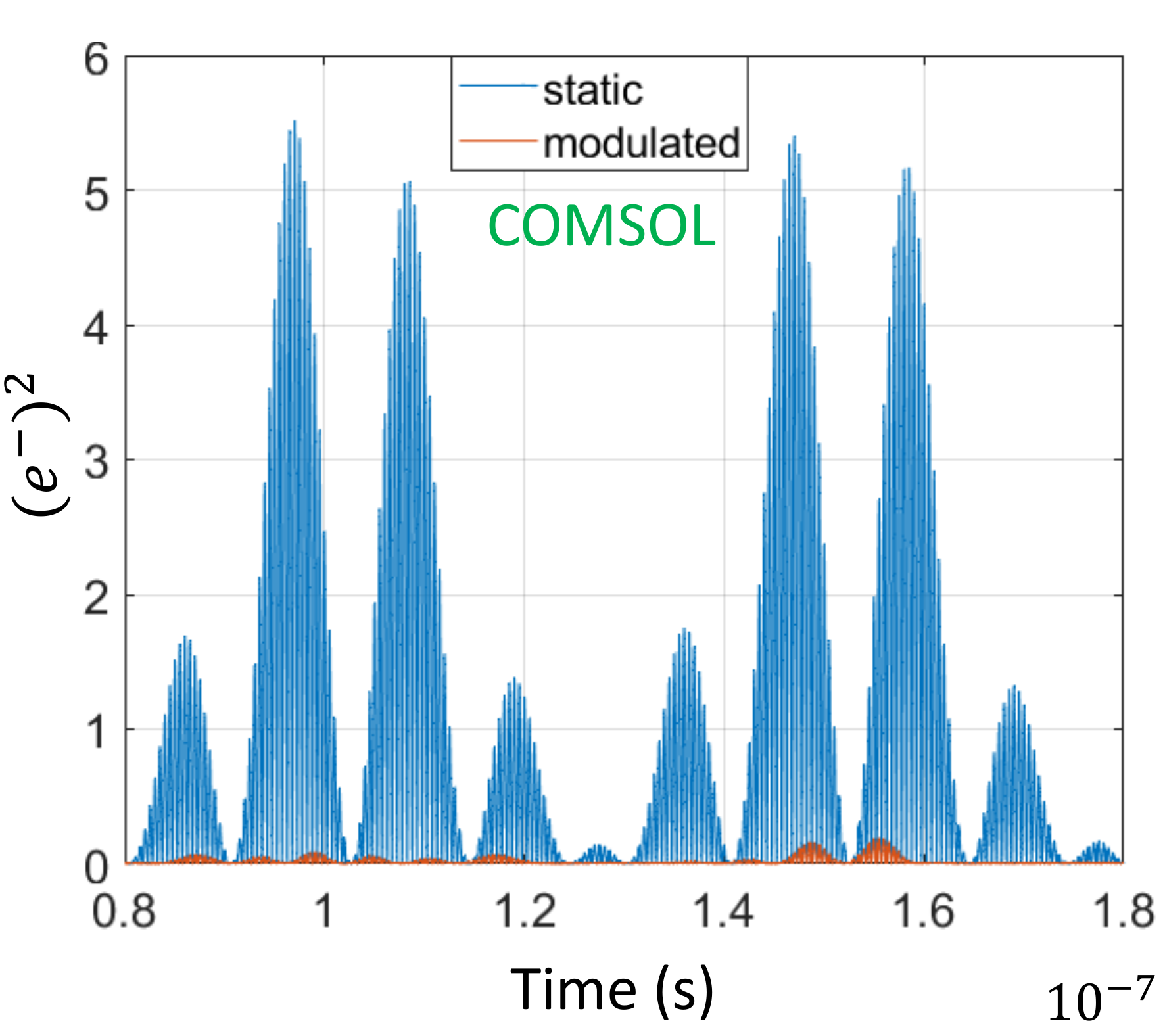}\label{fig:fig7b}}
\subfigure[]{\includegraphics[width=0.33\linewidth]{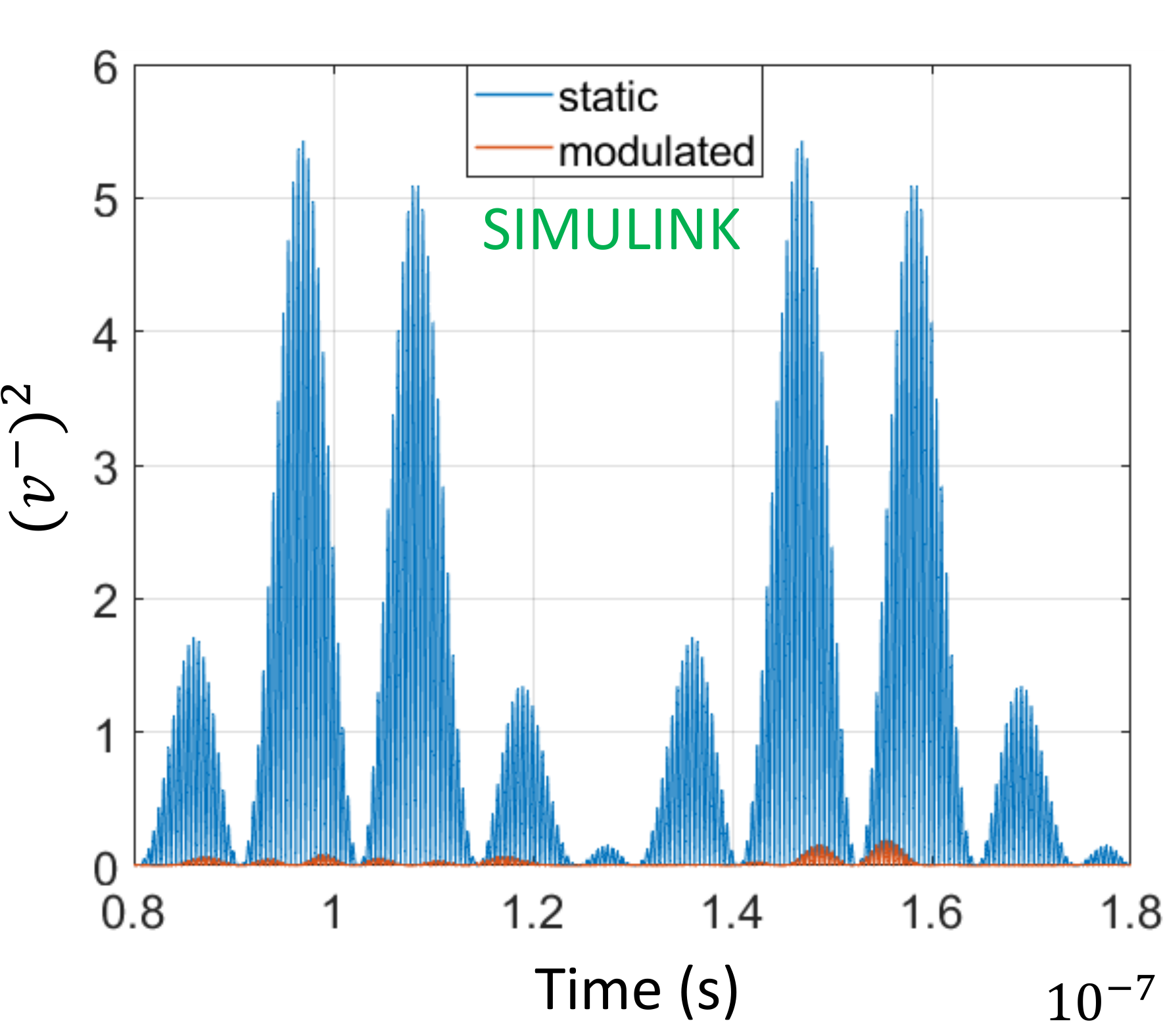}\label{fig:fi7c}}
\caption{Comparison between the reflected fields extracted from two different simulation tools. (a) The structure studied using COMSOL, the dielectric thickness has been adjusted to $d_{\rm c}=2.36$~mm from $d=2.367$~mm to tune the resonance frequency to 1~GHz. (b) Total reflected electric field ($e^-$) squared as a function of time extracted from COMSOL. (c) Total reflected voltage ($v^-$) squared as a function of time extracted from SIMULINK.}
 \label{fig_Fig7}
\end{figure*}

Next, we explore the possibility to perfectly match a higher number of frequencies to a narrow-band absorber. We consider an input signal formed by four harmonics at 0.95~GHz, 0.97~GHz, 1.03~GHz, and 1.05~GHz, while the control signal is at 1~GHz. Those harmonics give rise to modulation frequencies of $\omega_{\rm m 1}=2 \pi 30 \times 10^6$~rad/sec and $\omega_{{\rm m} 2}=2 \pi 50 \times 10^6$~rad/sec. An array of patches is designed to produce a resonance with the shorted transmission line at 1~GHz with $L_{\rm d}= 3$~nH, $C_{\rm m}=8.44$~pF, and $g_0=1/\eta_0$. The structure parameters corresponding to these values are $\varepsilon_d=12$, $D=\lambda_0/8=37.5$ mm, $\ell=51 \:\mu{\rm m}$, and $d=2.367 \: {\rm mm}\approx\lambda/125$. For simplicity, we assume $a_{\pm2}=a_{\pm 1}=1$~V. Perfect absorption of all harmonics is realized with  $\phi_{{\rm m} 1,2}=3\pi/2$, $a_{0}=10$~V, $m_1=0.245$, and $m_2=0.38$. Note that we can decrease $m_k$ even more by increasing $a_0$. 
Numerical results obtained from SIMULINK are shown in Fig.~\ref{fig_Fig6}. The simulation schematic is shown in  Fig.~\ref{fig:sim2}, where a transmission line with $\varepsilon_d=12$ and $d=2.367$~mm is used to model the dielectric layer. The time-modulated effective resistance and the absorption are shown in Fig.~\ref{fig:fig6a} and Fig.~\ref{fig:fig6b}, respectively. In the absence of modulation, the metasurface provides narrow-band absorption, however, when modulation is present, absorption is boosted nearly to perfection for all harmonics. For example, at 0.95~GHz, the absorption increases from approximately 50\% to 99\% overcoming the bandwidth limit of thin static linear absorbers (see Appendix~\ref{appa})~\cite{Rozanov}. At the same time, the modulation does not reduce the absorption at the control frequency, as the modulation products coupling to this frequency cancel out due to the antisymmetry. The reflected voltage amplitude squared as a function of frequency is shown in Fig.~\ref{fig:fig6d}, where we can see that it is negligible in case of modulation. The spurious harmonics are also negligible due to the large amplitude of the control signal $a_0$. It is important to stress that the incident power is perfectly absorbed in the resistive layer, and total reflected power is negligible over the whole frequency spectrum. In addition to boosting absorption to the maximum, it is also possible to tune the absorption to any desired value by varying $a_0$ and $\phi_{{\rm m} k}$. Figure~\ref{fig:fig6e} shows the absorption as a function of $a_0$ and $\phi_{{\rm m} k}$. For $\phi_{{\rm m} 1,2}=\frac{3 \pi}{2}$ (red dashed lines) the absorption is enhanced while $a_0 < 10$~V, and for $\phi_{{\rm m} 1,2}=\frac{ \pi}{2}$ (blue dashed lines) the absorption is decreased while $a_0 < 3.5$~V. In addition, different values can be assigned to $\phi_{{\rm m} 1}$ and $\phi_{{\rm m} 2}$, providing the ability to increase the absorption at some frequencies and decrease it at other frequencies. Figure~\ref{fig:fig6e} confirms that the absorption is fully tunablfe to any value between~0 and~1 by properly engineering the control signal and modulation parameters. In addition, it also shows that the system is insensitive to small changes in amplitudes that might arise from fabrication/synchronization imperfections. To further check the sensitivity of the proposed technique, Fig.~\ref{fig:fig6c} shows the absorption at frequency 0.95~GHz as a function of the modulation phase angle, which proves that the system is insensitive to small imperfections in phase-angle synchronization, hence, the system is stable and robust (see Appendix~\ref{appb}). Finally, to confirm the validity of the above numerical results, Fig.~\ref{fig_Fig7} shows a comparison between simulation results from SIMULINK and COMSOL. There is only one difference between both simulations, which is the thickness of the dielectric layer. The transmission line simulated in SIMULINK has the length $d=2.367$~mm, however, in COMSOL the thickness $d_{\rm c}=2.36$~mm has been used to compensate for a small shift of the resonance frequency due to parasitic reactive fields at the boundaries of the computation domain.   Figure~\ref{fig_Fig7} shows the total reflected fields squared as a function of time from both tools. In both cases, the static and dynamic, the total reflection matches well. This comparison confirms the validity of the SIMULINK circuit simulations shown in Fig.~\ref{fig_Fig6}.

It is important to stress that the metasurface does not pump energy to the fields, and the sum of the reflected and absorbed powers is equal to the incident power carried by all input waves. In addition, we note that although we have given examples of designs for the microwave range, this technique can be used at any other frequency band. Indeed,  the described effects do not depend on the actual value of the carrier frequency: the modulation is defined by the frequency difference between the control signal and the input signals. 

The theory presented here applies not only to flat and thin resonant absorbers, but to any lossy resonant system (e.g., resonant scatterers or antennas). In the vicinity of the resonant frequency, any such system can be modeled by a lossy resonant circuit shown in Fig.~\ref{cap}(c). The equivalent circuit parameters can be found as functions of the antenna or scatterer geometry and materials, see e.g.~\cite{Balanis}. Importantly, for antennas, the circuit parameters can be tuned and controlled by the antenna load. In particular, the circuit resistance is the sum of the radiation resistance, the loss resistance, and the resistance of the antenna load. The load resistance can be modulated in time using the same techniques as presented in this paper (see Appendix~\ref{appc}) for applications in thin absorbers. Thus, we see that using time-varying resistive loads of antennas, it is possible to reach the maximum level of absorption by resonant dipole antennas (e.g., \cite{plasmonics}) at several frequencies inside the resonant band,  instead of only a  single resonance frequency. 
Similar results can be also obtained for the fundamental resonance band of plasmonic nanoparticles~\cite{plasmonics,Benz:15} or Mie resonances in dielectric spheres, as these objects can be also modeled using the same circuit model. However, modulating the absorption level in these cases can be practically challenging.

	
\section{Conclusions}
\label{concl}

In this paper, the theory of temporally modulated passive lossy objects  illuminated by two or more waves having different frequencies was presented. In particular, we considered planar boundaries illuminated by plane waves. It was shown that waves reflected from  lossy boundaries modulated by signals that are properly synchronized with the spectral content of the incident radiation are functions of the modulation parameters and the amplitudes of the incident waves, although the system is governed by linear differential equations. The results have proven that by properly designing these factors, full control of the reflected waves is achieved. This opens many possibilities, such as realizing reflection similar to that from perfectly black or reactive boundaries. We have shown that  dynamic purely resistive boundaries are  equivalent to a static impedance boundary that has both static resistive and reactive components. An expression was derived for this virtual reactive component imposed by the modulation, and, also, the power relations were discussed.

This feature can be used 
in many applications, however, here, specifically, we used 
it to design multifrequency perfect absorbers. It was shown that by properly designing the modulation parameters and the amplitude of the control signal, it is possible to achieve unity absorption for multiple frequencies, even for ultranarrow-band absorbers, overcoming the bandwidth limit of passive static linear absorbers. In addition, it was shown that the absorption can be fully tuned remotely by tuning the control signal's amplitude. All these results are associated with a resistive layer that does not provide gain and is unconditionally stable. Furthermore, slow modulation was used instead of the practically limited fast modulation which is typically used in parametric systems. We stress that the theory provided in this paper is general and can be used in other electromagnetic/optical fields such as plasmonics or nanophotonics.

Finally, we note that illuminating waves do not have to be mutually coherent. For example, if we illuminate by two input waves and  their phase difference arbitrarily varies in time, similar effects can be achieved using non-monochromatic low-frequency modulation synchronized with this time-varying phase difference.

\bigskip
\section{Acknowledgments} 
This work was supported by the Academy of Finland under grants 330957 and 330260. This work was partially supported by the Spanish Ministerio de Educaci\'on y Formaci\'on Profesional under the  grant Beatriz Galindo BG20/00024. This work is based on the master's thesis of the first author Mohamed Mostafa. The authors wish to thank Dr.~Xuchen Wang for helping to draw Fig.~1 of this paper. The authors also wish to thank Dr.~Prasad Jayathurathnage and Dr.~Grigorii Ptitcyn for useful discussions. 


\appendix

\section{Advantages of the proposed time-varying absorber over a static absorber} \label{appa}

It is possible for a static linear  system to have multiple reflection zeros. Several reflection zeros can be realized in electrically thick absorbers, such as Jaumann absorbers, for example. However, these static systems have fundamental limitations and disadvantages compared to the proposed time-varying system. Here, we deal with electrically thin absorbing layers. As mentioned in the introduction, the bandwidth of a static linear absorber is limited by its thickness (the Rozanov limit for linear static Dällenbach screens). If thin absorbers have several resonant frequencies corresponding to several reflection zeros, then this bandwidth limit applies to every narrow absorption band in the vicinity of each resonance. Moreover, if a passive  thin absorber has several reflection zeros, it is inevitable that between each of these absorption peaks there are reflection maxima. Thus, it is in principle impossible to  bring them arbitrary close to each other and ``merge'' them. This property follows from the Foster reactance theorem: between absorption maxima, reactance must always grow. Thus, it must cross zero between every two poles. The proposed time-modulated structure overcomes these fundamental limitations and allows creation of multiple reflection zeros \emph{inside every resonant absorption peak}. Multifrequency perfect absorption is achieved while the thickness of the absorber remains extremely thin. These functionalities are not possible in passive linear systems. 

Another advantage of the proposed technique is that there is no theoretical limit for the number of harmonics that the proposed system can absorb perfectly. Two examples are shown for input signals with two and four harmonics, and perfect absorption was achieved for both. Furthermore, the same theory applies for higher numbers of harmonics, for example, 10 harmonics, with no added complexity. On the other hand, for a static linear system the number of zeros is limited. In static linear systems there is an optimization burden for achieving  desirable performance, even if the absorber is not compact.

In addition, for the proposed time-varying system, the position of input harmonics and the spacing between them (on one side of the carrier) can be arbitrary. For example, the case shown in Fig.~\ref{fig_Fig6} shows that on the right side of the carrier there are two harmonics at 1.03~GHz and 1.05~GHz. Let us also assume that there is another input harmonic at 1.02~GHz. By designing the modulation accordingly, it is possible to perfectly absorb all harmonics even though they are spaced differently and they are relatively close to each other. We consider amplitude modulated input signals as a practically relevant example.  However, the proposed system can also work for single-sided input signals, in this case the modulation products produced on the other side of the carrier need to be filtered out. 

Another advantage of the proposed system is its tunability. As illustrated by Fig.~\ref{fig:fig6e}, it is possible to tune the absorption level to any value, not only by changing the modulation phase angle, but also by changing the control signal's amplitude, which provides a possibility for remotely tuning the absorption level.

\begin{figure}[]
\centering
{\includegraphics[width=1\linewidth]{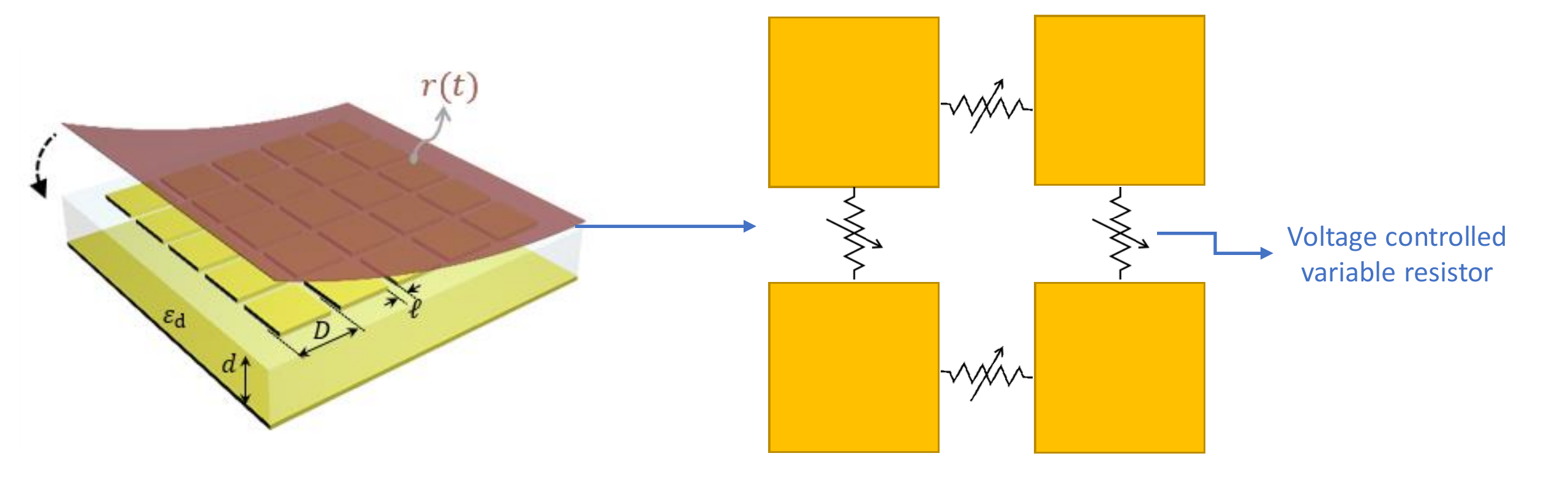}}
\caption{Possible practical realization of dynamic resistive metasurfaces.} 
\label{R1}
\end{figure}

\section{Sensitivity of the proposed technique to fabrication imperfections} \label{appb}

To achieve perfect absorption, conditions in Eq.~\ref{eq_coditions} should be met. These  conditions mainly define the carrier to harmonic amplitude ratio, the modulation amplitude, and the modulation phase angle. As a result, these are the critical parameters that affect the absorption in case of having imperfections in fabrication or synchronization issues in the modulating system.

Firstly, let us discuss how the system responds to changes in the amplitudes (carrier, harmonic or modulation amplitude). In Fig.~\ref{fig:fig6e}, it is shown how tuning the carrier amplitude (and also tuning the carrier to harmonic amplitude ratio) changes the absorption level. For example, when the carrier amplitude changes from 10 to 5~V, the absorption at frequency 0.95 GHz changes from 1 to 0.84. Hence, this shows that a slight change in the carrier to harmonic amplitude ratio (or the modulation amplitude) would not decrease the absorption level significantly.

Secondly, regarding the modulation phase angle, Fig.~\ref{fig:fig6c} shows how tuning the modulation phase angle affects the absorption level. It is clear that slight changes in the modulation phase angle do not affect the absorption level significantly.

As a result, the system is robust, stable, and nearly perfect absorption can be achieved even in the presence of some imperfections.

\section{On practical realizations} \label{appc}             
                   
Here, we discuss how the proposed system can be realized practically. A dispersionless thin resistive sheet at microwaves can be realized as a thin conductive (metal) sheet, as discussed e.g. in \cite{sergeiblue}, Section~2.5. The sheet impedance of thin metal foils is nearly purely resistive and equals $\frac{1}{\sigma d_{\rm{r}}}$,  where $\sigma$ is the metal conductivity, and $d_{\rm{r}}$ is the sheet thickness. At microwaves, metal conductivity is constant over wide frequency ranges. As a result, the sheet resistance practically does not depend on the frequency. We also see that in this case time modulation of sheet resistance can be achieved by modulating  conductivity of the layer material.

On the other hand, one practical realization for the proposed dynamic metasurface at microwave frequencies is shown in Fig.~\ref{R1}. Here, resistance modulation can be realized using electronic circuits. Thin metal patches positioned on a grounded substrate can be fabricated using conventional printed circuit board (PCB) techniques, which is a practical solution. The resistive sheet is partially shorted due to the presence of  metal patches, hence, the sheet resistance is defined by resistive components connected between the patches. 

\begin{figure}[]
\centering
\subfigure[]{\includegraphics[width=0.48\linewidth]{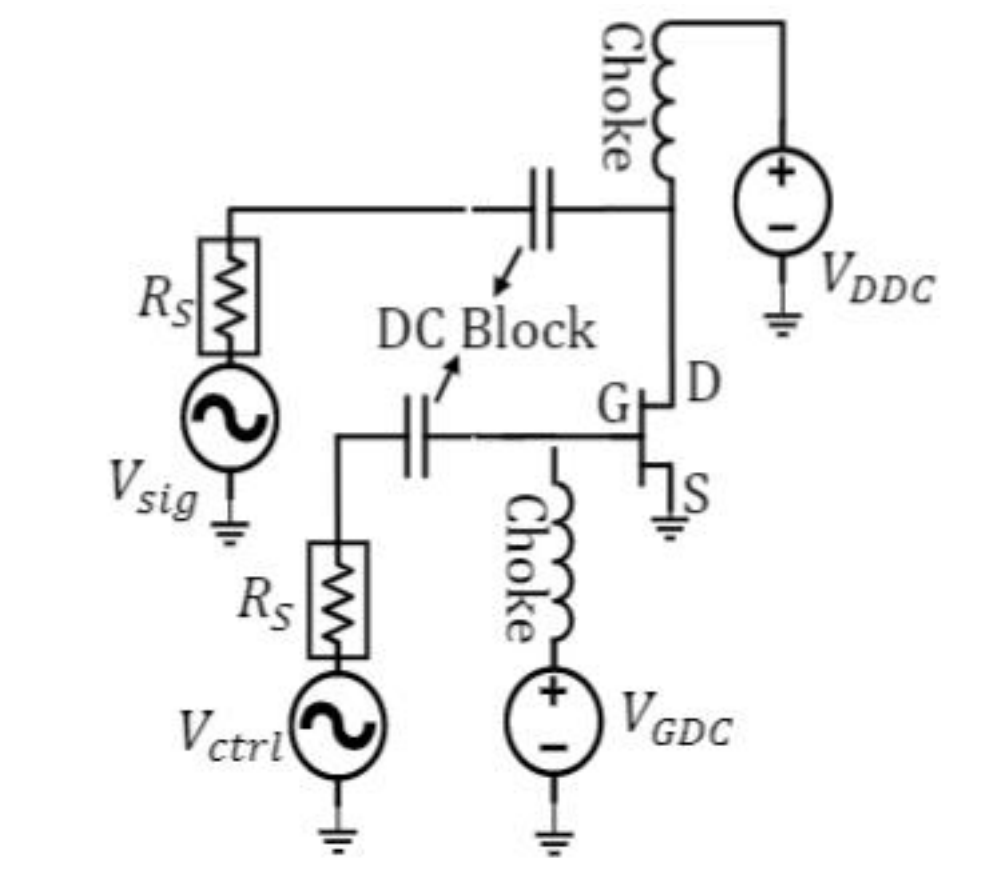}\label{}}
\subfigure[]{\includegraphics[width=0.48\linewidth]{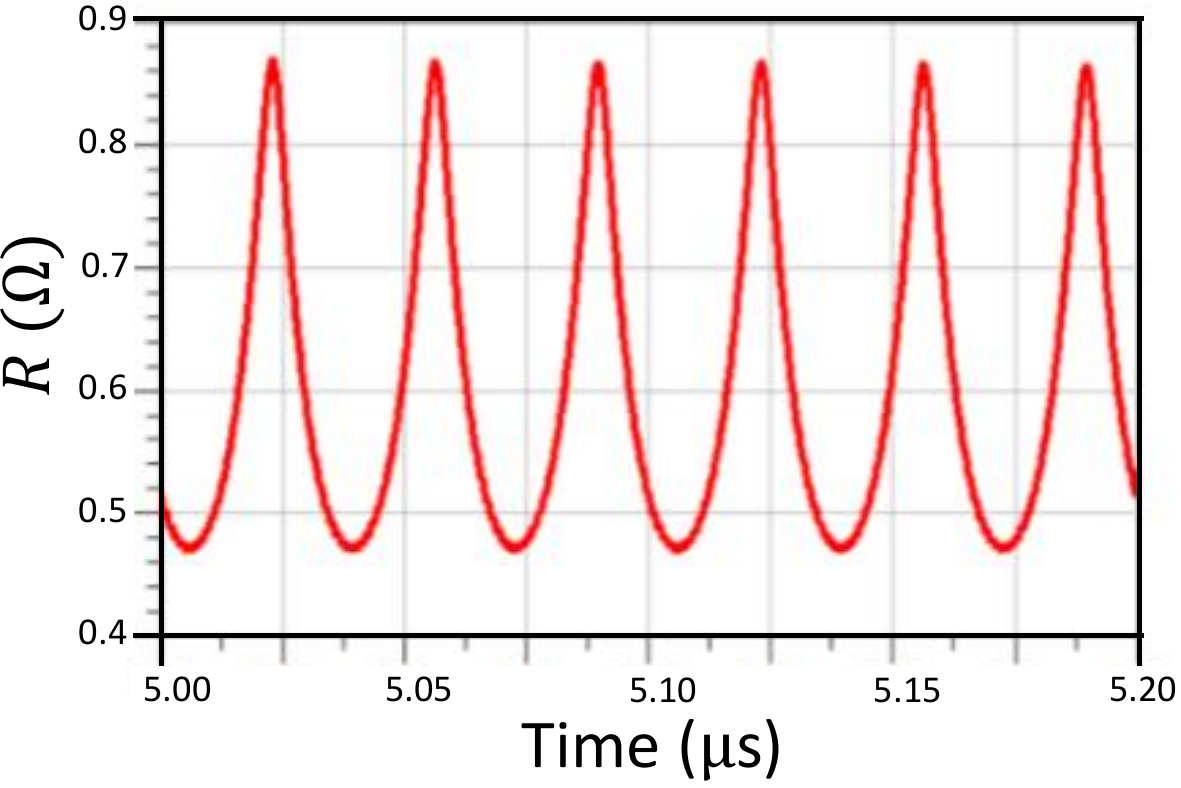}\label{}}
\caption{Time-modulated resistor design using a  MOSFET transistor. (a) Topology. (b) Simulated time-varying resistance as a function of time using Advanced Design System (ADS).}
\label{R2}
\end{figure}

Regarding practical realization of dispersionless time-varying resistive components, Fig.~\ref{R2} shows one proposal using a MOSFET transistor. By connecting a time-varying voltage source to the gate terminal, it is possible to obtain a voltage-controlled time-varying resistor that is practically  dispersionless within the operating frequency band (also see Appendix~\ref{appd}).


\begin{figure}[]
\centering
{\includegraphics[width=0.95\linewidth]{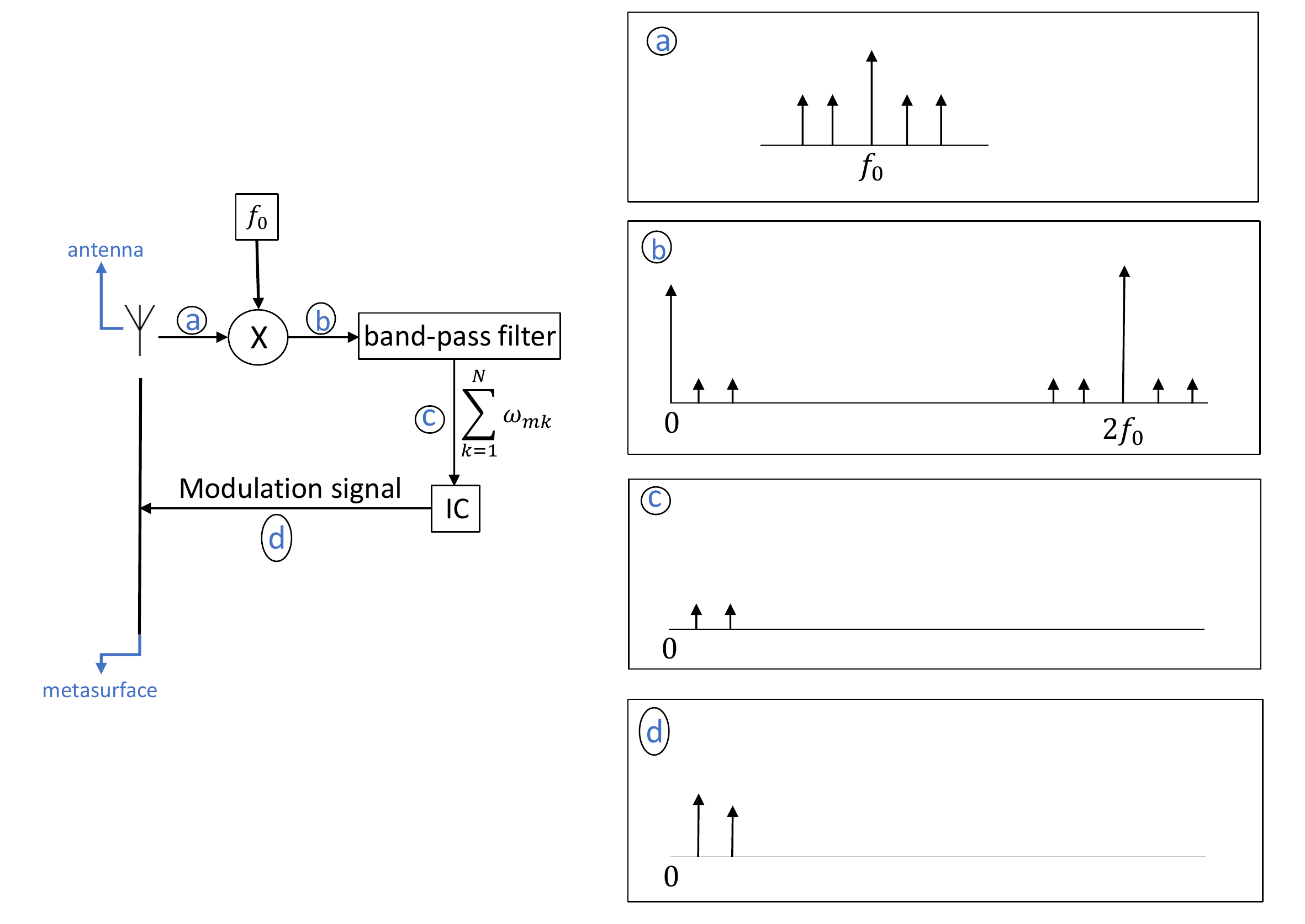}}
\caption{Schematic of a  proposed scheme for modulation signal synthesis and synchronization.} 
\label{R3}
\end{figure}

Finally, Fig.~\ref{R3} illustrates an example of a modulation scheme for creation and synchronization of  modulation signals. A sample of the incident field (a) is received by an antenna/sensor, then the incident signal is multiplied by the control signal, which results in spectrum (b). A band-pass filter is used to select the modulation frequencies as in (c). Then, an electronic chip is used to calculate the amplitudes and phases of the modulation harmonics. This process is similar to the demodulation process used in communication  systems. We note also that, as mentioned in Appendix~\ref{appb}, the system is not significantly sensitive to phase synchronization imperfections, which eases the complexity of the modulation scheme in Fig.~\ref{R3}.

\section{Power consumed to achieve modulation} \label{appd}

Let us assume that there is an input voltage source $V_{\rm{sig}}$ that should be connected to a time-varying resistor. As an example implementation of a time-modulated resistor we use a MOSFET transistor. As MOSFET transistors have the current-voltage relations of a resistor in the linear region, we obtain a time-varying resistor by applying some modulation signal to the transistor gate.  
A MOSFET transistor is connected to the voltage source $V_{\rm{sig}}$ as shown in Fig.~\ref{R2}(a). By configuring the DC voltage sources $V_{\rm{DDC}}$ and $V_{\rm{GDC}}$, and having $V_{\rm{ctrl}}$ turned off, the transistor operates in the linear regime and behaves as a static  resistor $R$. As a result, the input signal goes from $V_{\rm{sig}}$ to the ground and passes through the resistor $R$. One of the main characteristics of MOSFETs is having nearly zero gate current, meaning that practically no current can flow from/to gate to/from drain and source. Also, let us ignore the DC current induced by $V_{\rm{DDC}}$, as it can be filtered using a DC block.

To obtain a time-varying resistor, we turn $V_{\rm{ctrl}}$ on. By assigning a single harmonic signal to $V_{\rm{ctrl}}$ with a suitable amplitude level, the resistance of the transistor becomes time-varying $R(t)$, as shown in Fig.~\ref{R2}(b). Again, the input signal goes from $V_{\rm{sig}}$ to the ground passing by $R(t)$, which is what we want to obtain: a signal going through a time-varying resistor. Importantly, the active component of the gate current is very close to zero no matter what is the gate voltage, and 
the  power consumption of the modulation action is negligible. To make this point more clear, if $V_{\rm{sig}}$ is turned off, the active current going through all transistor terminals will be negligible, even if $V_{\rm{ctrl}}$ is on. Finally, this topology has been verified experimentally and we found that the power consumption due to the gate current is indeed negligible, as the gate current is small and mainly reactive. Also please note that the power consumed in the transistor's resistance $R$ in case of static and dynamic $R$ are different, but this power consumption is induced by the input signal, so it is not considered as the power consumed to achieve modulation.

\begin{figure}[]
\centering
{\includegraphics[width=0.95\linewidth]{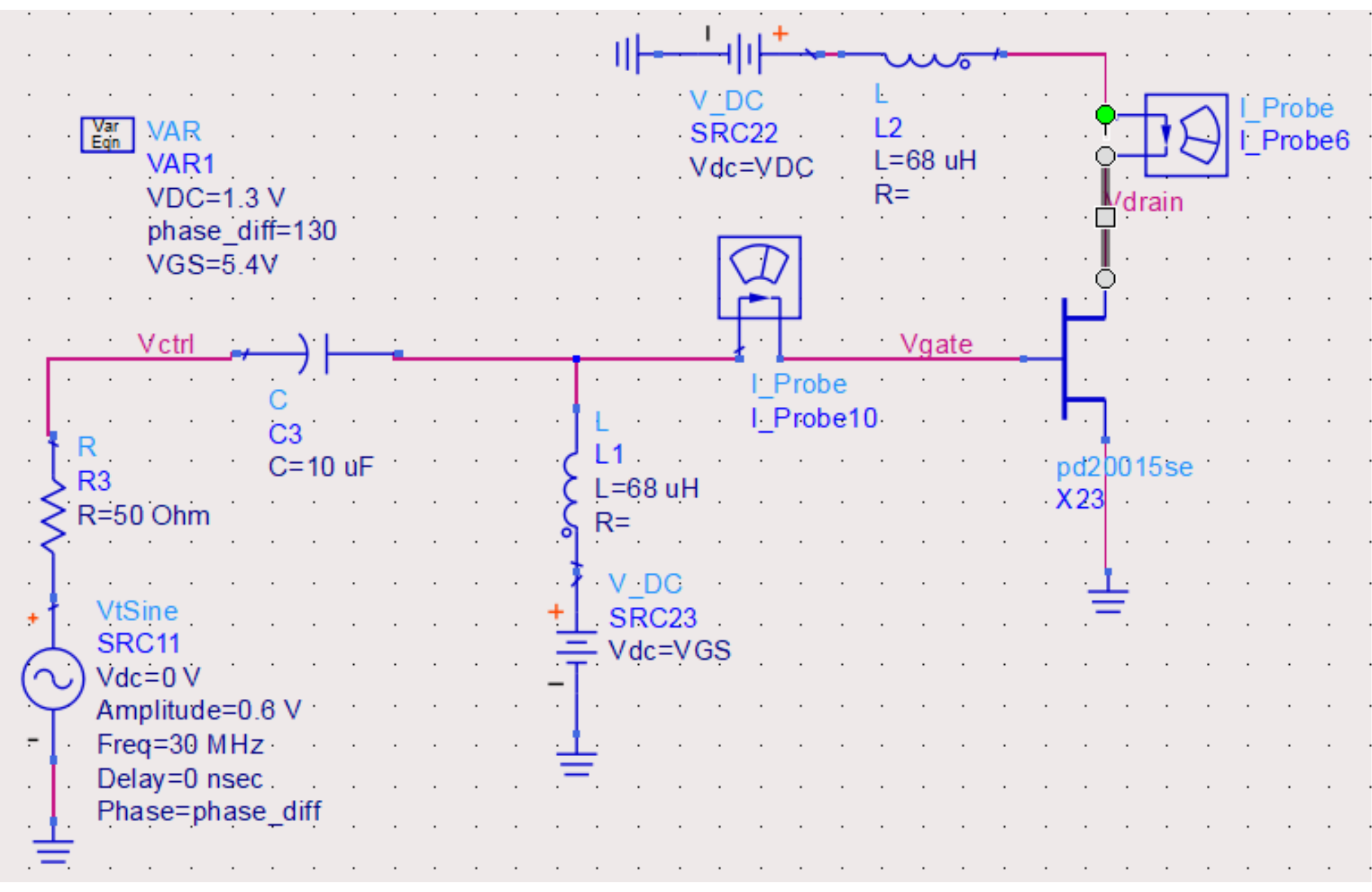}}
\caption{Circuit of time varying resistor using a particular MOSFET transistor.} 
\label{R4}
\end{figure}

\begin{figure*}[]
\centering
\subfigure[]{\includegraphics[width=0.40\linewidth]{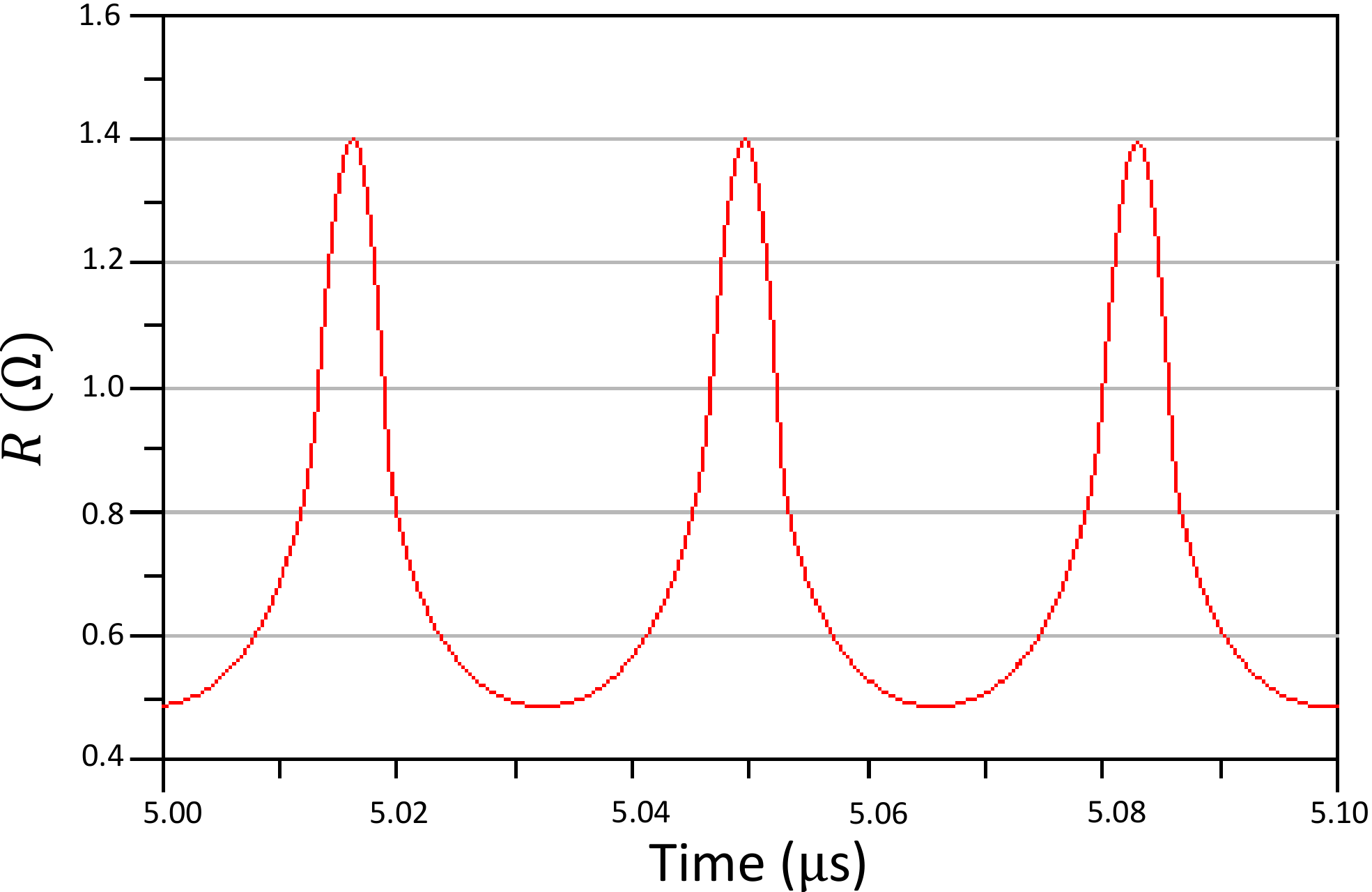}}
\hspace{2em}%
\subfigure[]{\includegraphics[width=0.40\linewidth]{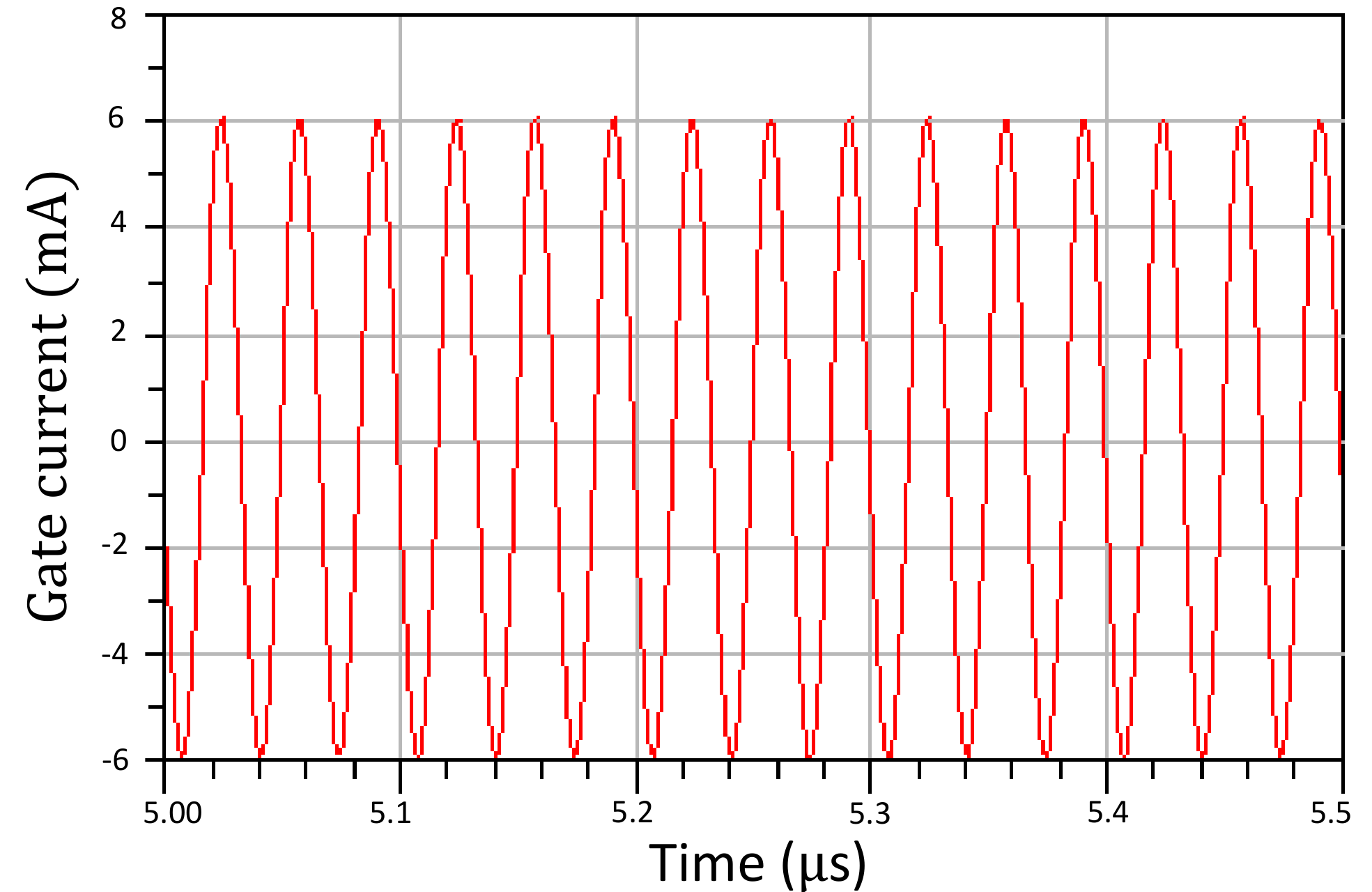}}
\subfigure[]{\includegraphics[width=0.4\linewidth]{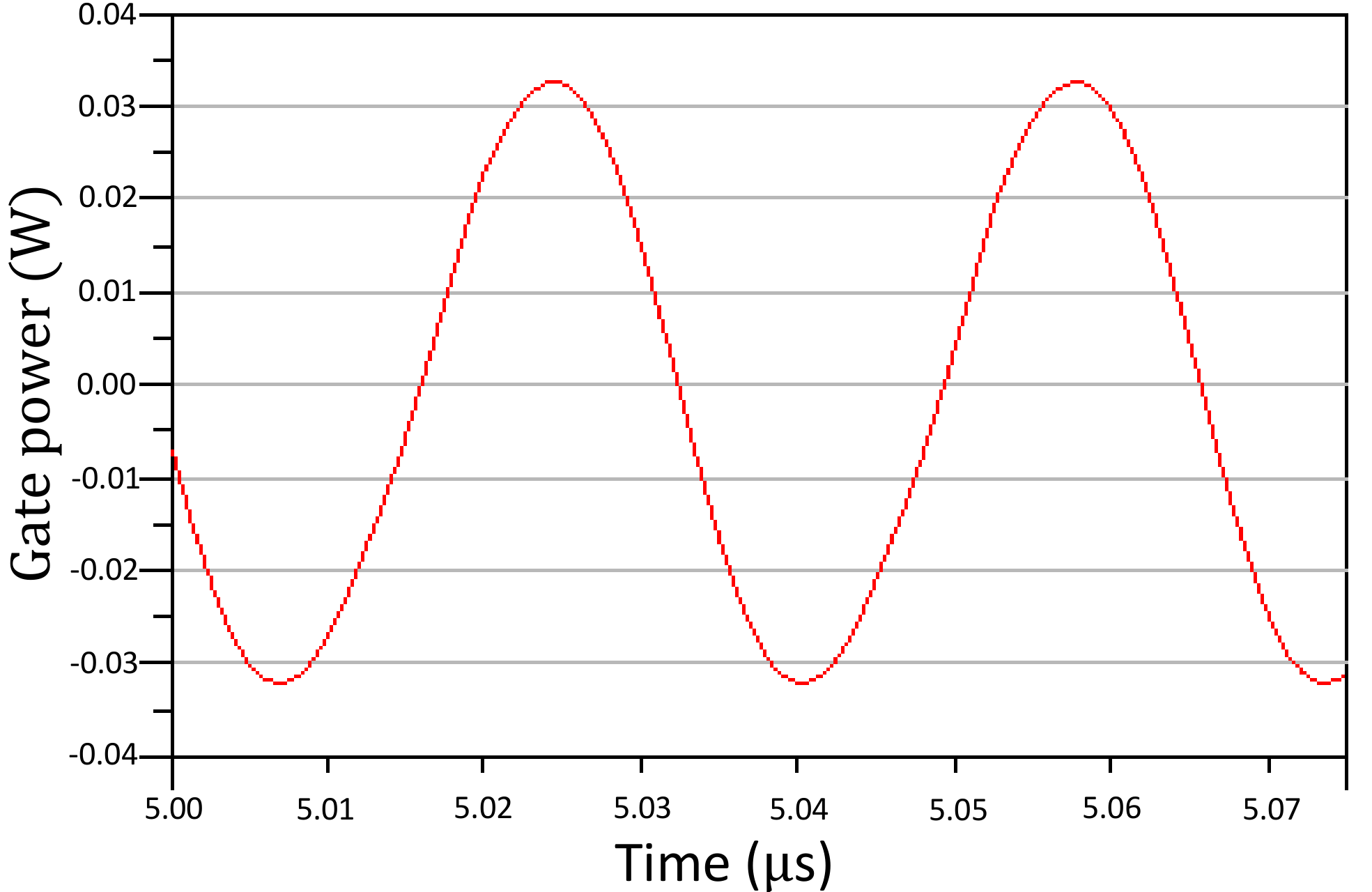}}
\hspace{1em}%
\subfigure[]{\includegraphics[width=0.45\linewidth]{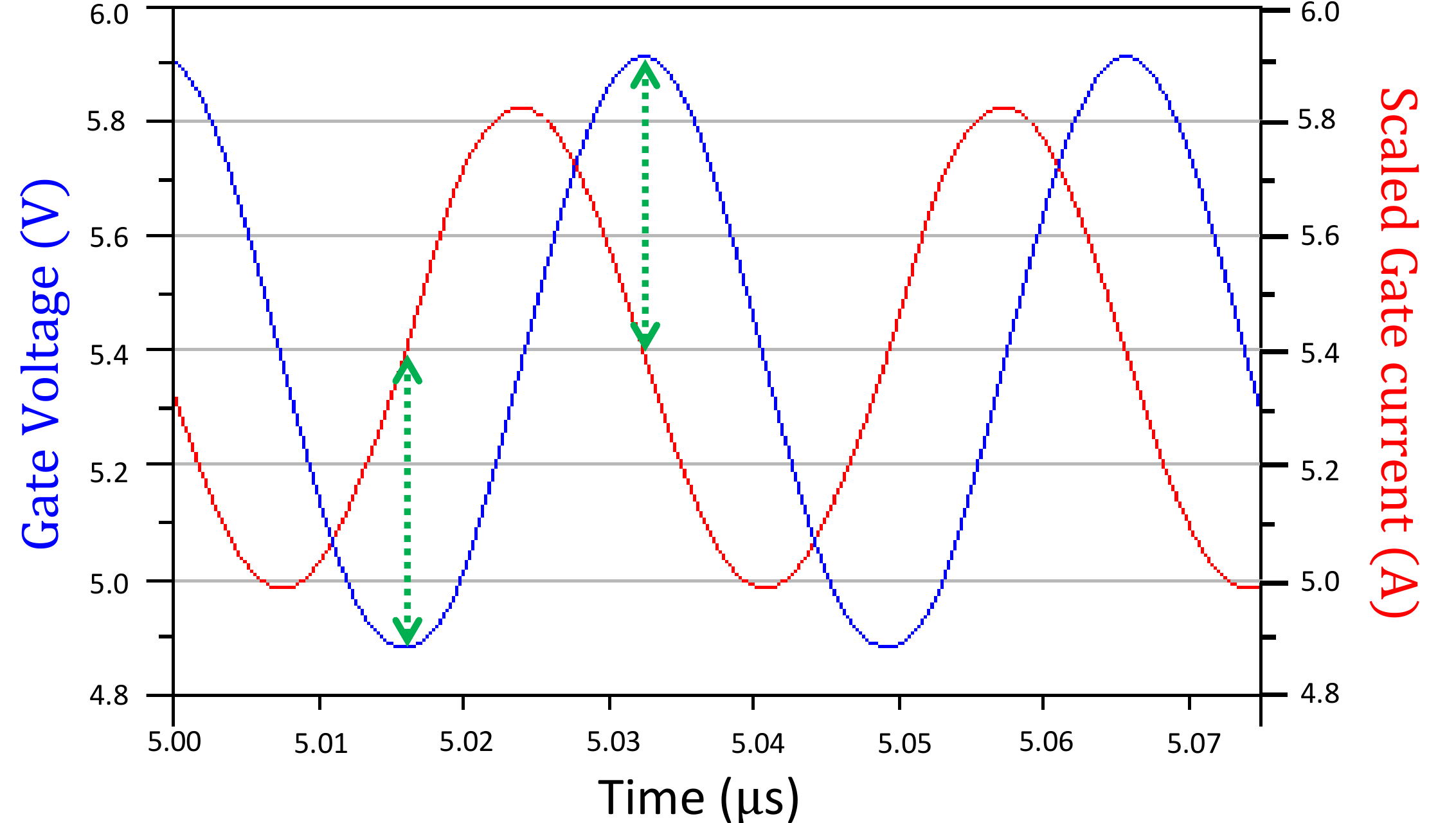}}
\caption{(a) The obtained time varying resistance. (b) Gate current. (c) Instantaneous gate power. (d) Gate voltage along with the gate current scaled and added to a DC level so that we can look at the phase difference.} 
\label{R5}
\end{figure*}

To provide an estimate of the consumed power, we verify the above conclusion through circuit simulations using ADS. Figure~\ref{R4} shows the circuit of the time-varying resistor, the circuit model for the PD20015-E MOSFET transistor that we use in experiments. The drain is connected to a 1.3 DC voltage source, the gate is connected to a 5.4 DC voltage source in addition to an AC voltage source with the amplitude 0.6 V at frequency 30 MHz, and the source is grounded. The obtained time-varying resistance is shown in Fig.~\ref{R5}(a), the gate current is shown in Fig.~\ref{R5}(b), and the instantaneous gate power is shown in Fig.~\ref{R5}(c). It is clear that the average current and power corresponding to losses are negligible, reading 0.3~$\mu$A and 5.5~$\mu$W,  respectively. Please note that the maximum power dissipation for this MOSFET is 79~W, hence, 5.5~$\mu$W is considered negligible compared to the power level at the time-varying resistor. To further confirm that the gate current and power are mainly reactive (this current flows through the gate capacitance and a very small parasitic resistance), Fig.~\ref{R5}(d) shows the gate voltage along with the gate current scaled and added to a DC level, so that we can look at the phase difference. It can be seen that the current is indeed leading the voltage by approximately 90 degrees, meaning that the gate power is approximately  reactive. Also note that from Fig.~\ref{R5}(b) we see that the DC component of the gate current is negligible, so that there is time-varying reactive current. Hence, turning on the time-harmonic voltage source connected to the gate does not result in noticeable power dissipation/loss. Again, the power going through the resistance $R$ is different when the resistance is time varying, but all the power dissipated in $R$ comes  from the drain terminal, while no active power is drawn from the harmonic source connected to the gate.

Based on the above observations, we believe that it is in principle possible to obtain a time-varying resistor without consuming any significant power. This is in agreement with Ohm's law. From the mathematical point of view, for reactive elements, there has to be power gain/loss when a reactive element is time-varying. This is because there is one term
added to the voltage-current relation due to the non-zero
capacitance/inductance time derivative, and this term indicates power gain/loss due to the exchange of power between the main circuit and the modulation circuit. However, Ohm's law remains the same even if the resistor is time-varying, indicating that there is no need to loose/gain power when a resistor is time-varying.

Regarding the modulating system responsible for modulation signal synchronization and synthesis, it will consume some power as any electronic device. However, conventional matching techniques that would
seek similar performance are much more complicated than the proposed system (see Fig.~\ref{R3}),
resulting in probably more power consumption.
Detailed calculation for the power consumed in the whole system depends heavily on particular
implementation, however, we strongly believe that proposed schemes like the one showed in
Fig.~\ref{R3} consume minimum power.

\bibliography{references}
	
\end{document}